%% file: main.tex
\pdfoutput=1

\documentclass[12pt,a4paper]{article}

\usepackage{ifthen} 
\newboolean{pdflatex}
\setboolean{pdflatex}{false} 

\newboolean{articletitles}
\setboolean{articletitles}{true} 

\newboolean{uprightparticles}
\setboolean{uprightparticles}{false} 

\newboolean{inbibliography}
\setboolean{inbibliography}{false} 

\usepackage{multirow}
\usepackage{tikz}
\usetikzlibrary{decorations.pathreplacing,decorations.pathmorphing,decorations.markings,trees,calc,patterns,snakes,arrows}
\tikzset{
photon/.style={decorate, decoration={snake}, draw=red},
particle/.style={draw=blue, postaction={decorate},decoration={markings,mark=at position .5 with {\arrow[draw=blue,scale=2]{>}}}},
antiparticle/.style={draw=blue, postaction={decorate},decoration={markings,mark=at position .5 with {\arrow[draw=blue,scale=2]{<}}}}, 
gluon/.style={decorate, draw=black,decoration={snake,amplitude=4pt, segment length=5pt}}, 
majorana/.style={draw=black, postaction={decorate},decoration={markings,mark=at position .48 with {\arrow[draw=black]{>}},mark=at position .52 with {\arrow[draw=black]{<}}}},
gluonloop/.style={circle, decorate, draw=black, decoration={coil,aspect=1.2,amplitude=2pt, segment length=4pt},minimum height=1.2em},
}

\input{preamble}
\usepackage{longtable} 

\begin{document}

\renewcommand{\thefootnote}{\fnsymbol{footnote}}
\setcounter{footnote}{1}

\input{title-LHCb-PAPER}


\renewcommand{\thefootnote}{\arabic{footnote}}
\setcounter{footnote}{0}



\pagestyle{plain} 
\setcounter{page}{1}
\pagenumbering{arabic}


\input{introduction}

\input{detector}

\input{selection}

\input{fit}

\input{systematics}

\input{results}

\input{acknowledgements}

\addcontentsline{toc}{section}{References}
\setboolean{inbibliography}{true}
\bibliographystyle{LHCb}
\bibliography{main,dsphi,LHCb-DP,LHCb-PAPER}

\newpage

 
\newpage
\input{LHCb_Authorship_flat_02-Aug-2017.tex}

\end{document}

%% file: preamble.tex
\usepackage[top=1in, bottom=1.25in, left=1in, right=1in]{geometry}

\usepackage{microtype}
\usepackage{lineno}  
\usepackage{xspace} 
\usepackage{caption,subcaption} 

\usepackage{graphicx}  
\usepackage{color}
\usepackage{colortbl}
\graphicspath{{./figs/}} 

\usepackage{amsmath} 
\usepackage{amssymb}
\usepackage{amsfonts}
\usepackage{upgreek} 

\newcommand*\patchAmsMathEnvironmentForLineno[1]{%
\expandafter\let\csname old#1\expandafter\endcsname\csname #1\endcsname
\expandafter\let\csname oldend#1\expandafter\endcsname\csname
end#1\endcsname
 \renewenvironment{#1}%
   {\linenomath\csname old#1\endcsname}%
   {\csname oldend#1\endcsname\endlinenomath}%
}
\newcommand*\patchBothAmsMathEnvironmentsForLineno[1]{%
  \patchAmsMathEnvironmentForLineno{#1}%
  \patchAmsMathEnvironmentForLineno{#1*}%
}
\AtBeginDocument{%
\patchBothAmsMathEnvironmentsForLineno{equation}%
\patchBothAmsMathEnvironmentsForLineno{align}%
\patchBothAmsMathEnvironmentsForLineno{flalign}%
\patchBothAmsMathEnvironmentsForLineno{alignat}%
\patchBothAmsMathEnvironmentsForLineno{gather}%
\patchBothAmsMathEnvironmentsForLineno{multline}%
\patchBothAmsMathEnvironmentsForLineno{eqnarray}%
}

\usepackage{hyperxmp}

\usepackage{hyperref}    
\usepackage[all]{hypcap} 

\input{lhcb-symbols-def} 

\usepackage{cite} 
\usepackage{mciteplus}

%% file: lhcb-symbols-def.tex
\usepackage{xspace} 
\usepackage{upgreek}

\def\lhcb {\mbox{LHCb}\xspace}

\def\MagUp {\mbox{\em Mag\kern -0.05em Up}\xspace}

\ifthenelse{\boolean{uprightparticles}}%
{

 \def\Ppi         {\ensuremath{\uppi}\xspace}

 \def\Pphi        {\ensuremath{\upphi}\xspace}

 \def\Ppsi        {\ensuremath{\uppsi}\xspace}

 \def\PDelta      {\ensuremath{\Delta}\xspace}                 
 \def\PXi      {\ensuremath{\Xi}\xspace}                 
 \def\PLambda      {\ensuremath{\Lambda}\xspace}                 
 \def\PSigma      {\ensuremath{\Sigma}\xspace}                 
 \def\POmega      {\ensuremath{\Omega}\xspace}                 
 \def\PUpsilon      {\ensuremath{\Upsilon}\xspace}                 
 
 \def\PB      {\ensuremath{\mathrm{B}}\xspace}                 
                  
 \def\PD      {\ensuremath{\mathrm{D}}\xspace}

 \def\PJ      {\ensuremath{\mathrm{J}}\xspace}                 
 \def\PK      {\ensuremath{\mathrm{K}}\xspace}

 \def\PW      {\ensuremath{\mathrm{W}}\xspace}

 \def\Pb      {\ensuremath{\mathrm{b}}\xspace}                 
 \def\Pc      {\ensuremath{\mathrm{c}}\xspace}

 \def\Pi      {\ensuremath{\mathrm{i}}\xspace}

 \def\Pp      {\ensuremath{\mathrm{p}}\xspace}

 \def\Ps      {\ensuremath{\mathrm{s}}\xspace}                 
                  
 \def\Pu      {\ensuremath{\mathrm{u}}\xspace}

}
{

 \def\Ppi         {\ensuremath{\pi}\xspace}

 \def\Pphi        {\ensuremath{\phi}\xspace}

 \def\Ppsi        {\ensuremath{\psi}\xspace}                 
                  
 \mathchardef\PDelta="7101
 \mathchardef\PXi="7104
 \mathchardef\PLambda="7103
 \mathchardef\PSigma="7106
 \mathchardef\POmega="710A
 \mathchardef\PUpsilon="7107
                  
 \def\PB      {\ensuremath{B}\xspace}                 
                  
 \def\PD      {\ensuremath{D}\xspace}

 \def\PJ      {\ensuremath{J}\xspace}                 
 \def\PK      {\ensuremath{K}\xspace}

 \def\PW      {\ensuremath{W}\xspace}

 \def\Pb      {\ensuremath{b}\xspace}                 
 \def\Pc      {\ensuremath{c}\xspace}

 \def\Pi      {\ensuremath{i}\xspace}

 \def\Pp      {\ensuremath{p}\xspace}

 \def\Ps      {\ensuremath{s}\xspace}                 
                  
 \def\Pu      {\ensuremath{u}\xspace}

}

\makeatletter
\ifcase \@ptsize \relax
  \newcommand{\miniscule}{\@setfontsize\miniscule{4}{5}}
\or
  \newcommand{\miniscule}{\@setfontsize\miniscule{5}{6}}
\or
  \newcommand{\miniscule}{\@setfontsize\miniscule{5}{6}}
\fi
\makeatother

\DeclareRobustCommand{\optbar}[1]{\shortstack{{\miniscule (\rule[.5ex]{1.25em}{.18mm})}
  \\ [-.7ex] $#1$}}

\def\Wp     {{\ensuremath{\PW^+}}\xspace}

\def\uquark    {{\ensuremath{\Pu}}\xspace}
\def\uquarkbar {{\ensuremath{\overline \uquark}}\xspace}

\def\squark    {{\ensuremath{\Ps}}\xspace}

\def\cquark    {{\ensuremath{\Pc}}\xspace}

\def\bquark    {{\ensuremath{\Pb}}\xspace}
\def\bquarkbar {{\ensuremath{\overline \bquark}}\xspace}

\def\pion   {{\ensuremath{\Ppi}}\xspace}
\def\piz    {{\ensuremath{\pion^0}}\xspace}

\def\pip    {{\ensuremath{\pion^+}}\xspace}
\def\pim    {{\ensuremath{\pion^-}}\xspace}

\def\kaon    {{\ensuremath{\PK}}\xspace}
  \def\Kbar    {{\kern 0.2em\overline{\kern -0.2em \PK}{}}\xspace}

\def\KorKbar    {\kern 0.18em\optbar{\kern -0.18em K}{}\xspace}

\def\Kp      {{\ensuremath{\kaon^+}}\xspace}
\def\Km      {{\ensuremath{\kaon^-}}\xspace}

\def\Kstarz  {{\ensuremath{\kaon^{*0}}}\xspace}

\newcommand{\phiz}{\ensuremath{\Pphi}\xspace}

  \def\Dbar    {{\kern 0.2em\overline{\kern -0.2em \PD}{}}\xspace}
\def\D       {{\ensuremath{\PD}}\xspace}

\def\DorDbar    {\kern 0.18em\optbar{\kern -0.18em D}{}\xspace}
\def\Dz      {{\ensuremath{\D^0}}\xspace}
\def\Dzb     {{\ensuremath{\Dbar{}^0}}\xspace}
\def\Dp      {{\ensuremath{\D^+}}\xspace}
\def\Dm      {{\ensuremath{\D^-}}\xspace}

\def\Dstarzb {{\ensuremath{\Dbar{}^{*0}}}\xspace}
\def\Dstarp  {{\ensuremath{\D^{*+}}}\xspace}

\def\Dsp     {{\ensuremath{\D^+_\squark}}\xspace}
\def\Dsm     {{\ensuremath{\D^-_\squark}}\xspace}

\def\Dss     {{\ensuremath{\D^{*+}_\squark}}\xspace}
\def\Dssp    {{\ensuremath{\D^{*+}_\squark}}\xspace}

\def\B       {{\ensuremath{\PB}}\xspace}
\def\Bbar    {{\ensuremath{\kern 0.18em\overline{\kern -0.18em \PB}{}}}\xspace}

\def\BorBbar    {\kern 0.18em\optbar{\kern -0.18em B}{}\xspace}

\def\Bu      {{\ensuremath{\B^+}}\xspace}

\def\Bp      {{\ensuremath{\Bu}}\xspace}

\def\Bd      {{\ensuremath{\B^0}}\xspace}
\def\Bs      {{\ensuremath{\B^0_\squark}}\xspace}
\def\Bsb     {{\ensuremath{\Bbar{}^0_\squark}}\xspace}

\def\jpsi     {{\ensuremath{{\PJ\mskip -3mu/\mskip -2mu\Ppsi\mskip 2mu}}}\xspace}

  \def\Y#1S{\ensuremath{\PUpsilon{(#1S)}}\xspace}

\def\proton      {{\ensuremath{\Pp}}\xspace}

\def\Lz          {{\ensuremath{\PLambda}}\xspace}
\def\Lbar        {{\ensuremath{\kern 0.1em\overline{\kern -0.1em\PLambda}}}\xspace}
\def\LorLbar    {\kern 0.18em\optbar{\kern -0.18em \PLambda}{}\xspace}

\def\Lc      {{\ensuremath{\Lz^+_\cquark}}\xspace}

\newcommand{\decay}[2]{\ensuremath{#1\!\to #2}\xspace}         

\def\to                 {\ensuremath{\rightarrow}\xspace}

\def\CP                {{\ensuremath{C\!P}}\xspace}

\def\Vub  {{\ensuremath{V_{\uquark\bquark}}}\xspace}

\def\AT#1     {\ensuremath{A_{\mathrm{T}}^{#1}}\xspace}           

\def\C#1      {\ensuremath{\mathcal{C}_{#1}}\xspace}                       
\def\Cp#1     {\ensuremath{\mathcal{C}_{#1}^{'}}\xspace}                    
\def\Ceff#1   {\ensuremath{\mathcal{C}_{#1}^{\mathrm{(eff)}}}\xspace}        
\def\Cpeff#1  {\ensuremath{\mathcal{C}_{#1}^{'\mathrm{(eff)}}}\xspace}       
\def\Ope#1    {\ensuremath{\mathcal{O}_{#1}}\xspace}                       
\def\Opep#1   {\ensuremath{\mathcal{O}_{#1}^{'}}\xspace}                    

\newcommand{\tev}{\ifthenelse{\boolean{inbibliography}}{\ensuremath{~T\kern -0.05em eV}}{\ensuremath{\mathrm{\,Te\kern -0.1em V}}}\xspace}
\newcommand{\gev}{\ensuremath{\mathrm{\,Ge\kern -0.1em V}}\xspace}
\newcommand{\mev}{\ensuremath{\mathrm{\,Me\kern -0.1em V}}\xspace}
\newcommand{\kev}{\ensuremath{\mathrm{\,ke\kern -0.1em V}}\xspace}
\newcommand{\ev}{\ensuremath{\mathrm{\,e\kern -0.1em V}}\xspace}
\newcommand{\gevc}{\ensuremath{{\mathrm{\,Ge\kern -0.1em V\!/}c}}\xspace}
\newcommand{\mevc}{\ensuremath{{\mathrm{\,Me\kern -0.1em V\!/}c}}\xspace}
\newcommand{\gevcc}{\ensuremath{{\mathrm{\,Ge\kern -0.1em V\!/}c^2}}\xspace}
\newcommand{\gevgevcccc}{\ensuremath{{\mathrm{\,Ge\kern -0.1em V^2\!/}c^4}}\xspace}
\newcommand{\mevcc}{\ensuremath{{\mathrm{\,Me\kern -0.1em V\!/}c^2}}\xspace}

\def\mum  {\ensuremath{{\,\upmu\mathrm{m}}}\xspace}

\def\invfb   {\ensuremath{\mbox{\,fb}^{-1}}\xspace}

\newcommand{\chisq}{\ensuremath{\chi^2}\xspace}

\newcommand{\chisqip}{\ensuremath{\chi^2_{\text{IP}}}\xspace}

\def\gsim{{~\raise.15em\hbox{$>$}\kern-.85em
          \lower.35em\hbox{$\sim$}~}\xspace}
\def\lsim{{~\raise.15em\hbox{$<$}\kern-.85em
          \lower.35em\hbox{$\sim$}~}\xspace}

\def\sPlot{\mbox{\em sPlot}\xspace}

\def\ptot       {\mbox{$p$}\xspace}
\def\pt         {\mbox{$p_{\mathrm{ T}}$}\xspace}

\def\evtgen     {\mbox{\textsc{EvtGen}}\xspace}

\def\geant      {\mbox{\textsc{Geant4}}\xspace}

\def\photos     {\mbox{\textsc{Photos}}\xspace}

\def\pythia     {\mbox{\textsc{Pythia}}\xspace}

\def\tell1  {TELL1\xspace}
\def\ukl1   {UKL1\xspace}



%% file: title-LHCb-PAPER.tex

\begin{titlepage}
\pagenumbering{roman}

\vspace*{-1.5cm}
\centerline{\large EUROPEAN ORGANIZATION FOR NUCLEAR RESEARCH (CERN)}
\vspace*{1.5cm}
\noindent
\begin{tabular*}{\linewidth}{lc@{\extracolsep{\fill}}r@{\extracolsep{0pt}}}
\ifthenelse{\boolean{pdflatex}}
{\vspace*{-2.7cm}\mbox{\!\!\!\includegraphics[width=.14\textwidth]{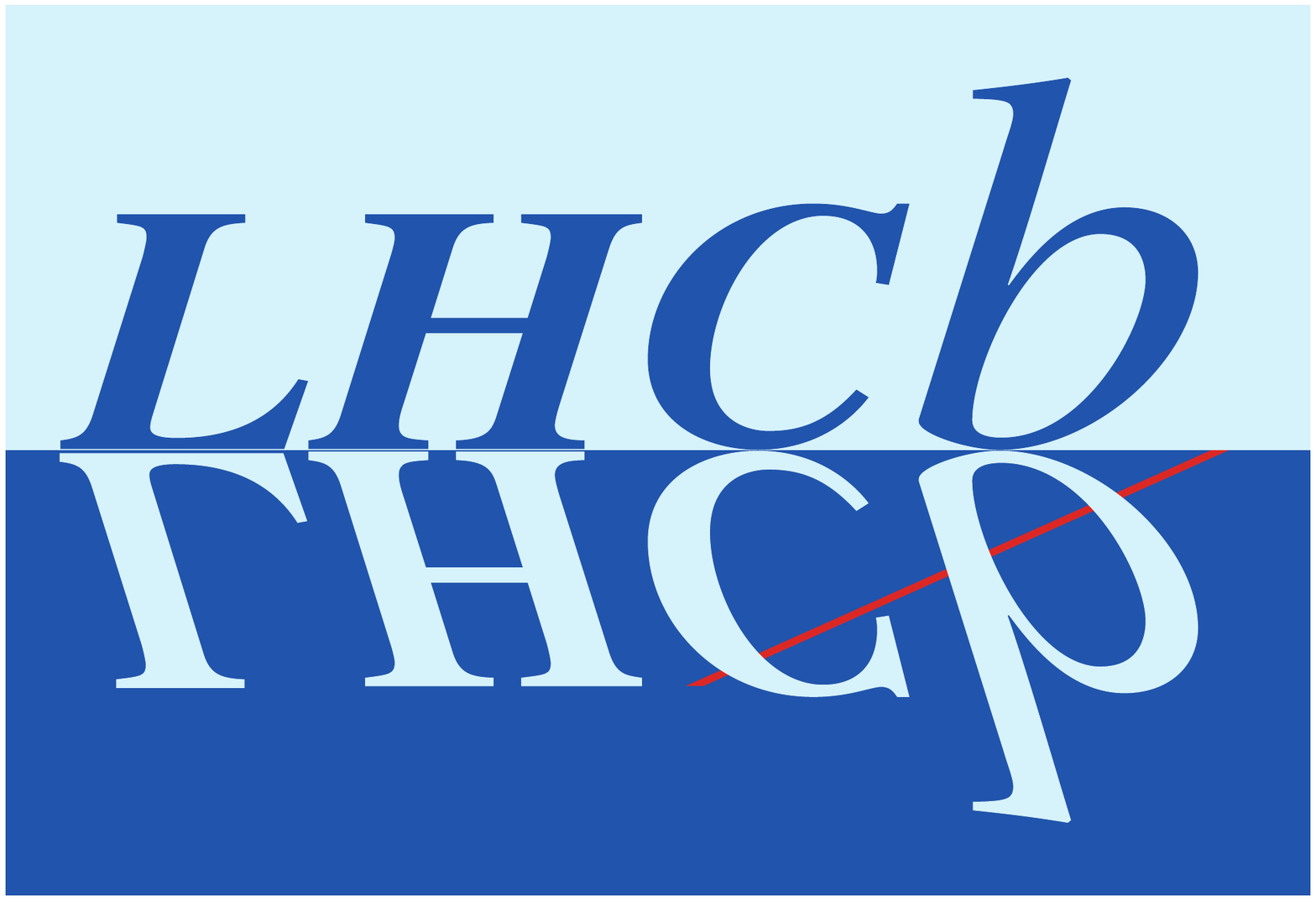}} & &}%
{\vspace*{-1.2cm}\mbox{\!\!\!\includegraphics[width=.12\textwidth]{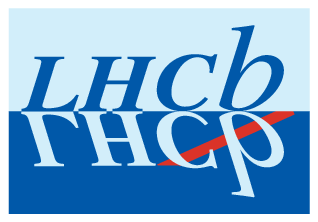}} & &}%
\\
 & & CERN-EP-2017-289 \\  
 & & LHCb-PAPER-2017-032 \\  
 & & November 15, 2017 \\ 
 & & \\
\end{tabular*}

\vspace*{4.0cm}

{\normalfont\bfseries\boldmath\huge
\begin{center}
  First observation of $B^{+} \to D_s^{+}K^{+}K^{-}$ decays and a search for $B^{+} \to D_s^{+}\phi$ decays
\end{center}
}

\vspace*{1.0cm}

\begin{center}
The LHCb collaboration\footnote{Authors are listed at the end of this paper.}
\end{center}

\vspace{\fill}

\begin{abstract}  

  \noindent
A search for $B^{+} \to D_s^{+}K^{+}K^{-}$ decays is performed using $pp$ collision data corresponding to an integrated luminosity of 4.8 fb$^{-1}$, 
collected at centre-of-mass energies of 7, 8 and 13\tev with the LHCb experiment. A significant signal is observed for the first time and the branching fraction is determined to be
\begin{equation*}
\mathcal{B}(B^{+} \to D_s^{+}K^{+}K^{-} ) = (7.1 \pm 0.5 \pm 0.6 \pm 0.7) \times 10^{-6}, 
\end{equation*}
\noindent where the first uncertainty is statistical, the second systematic and the third due to the uncertainty on 
the branching fraction of the normalisation mode $B^{+} \to D_s^{+} \overline{\kern -0.2em D}^{0}$.
A search is also performed for the pure annihilation decay $B^{+} \to D_s^{+}\phi$.
No significant signal is observed and a limit of
\begin{equation*}
\mathcal{B}(B^{+} \to D_s^{+}\phi) < 4.9 \times 10^{-7}~(4.2 \times 10^{-7})
\end{equation*}
is set on the branching fraction at 95\% (90\%) confidence level.

\end{abstract}

\vspace*{1.0cm}

\begin{center} 
  Published in JHEP 01 (2018) 131 
\end{center}

\vspace{\fill}

{\footnotesize 
\centerline{\copyright~CERN on behalf of the \lhcb collaboration, licence \href{http://creativecommons.org/licenses/by/4.0/}{CC-BY-4.0}.}}
\vspace*{2mm}

\end{titlepage}


\newpage
\setcounter{page}{2}
\mbox{~}
%
%
%
%

\cleardoublepage

%% file: introduction.tex

\section{Introduction}
\label{sec:Introduction}

\begin{figure}[!h]
    \centering 
        \includegraphics[trim={0.5cm 1.0cm 0.5cm 0.5cm},width=1.0\textwidth]{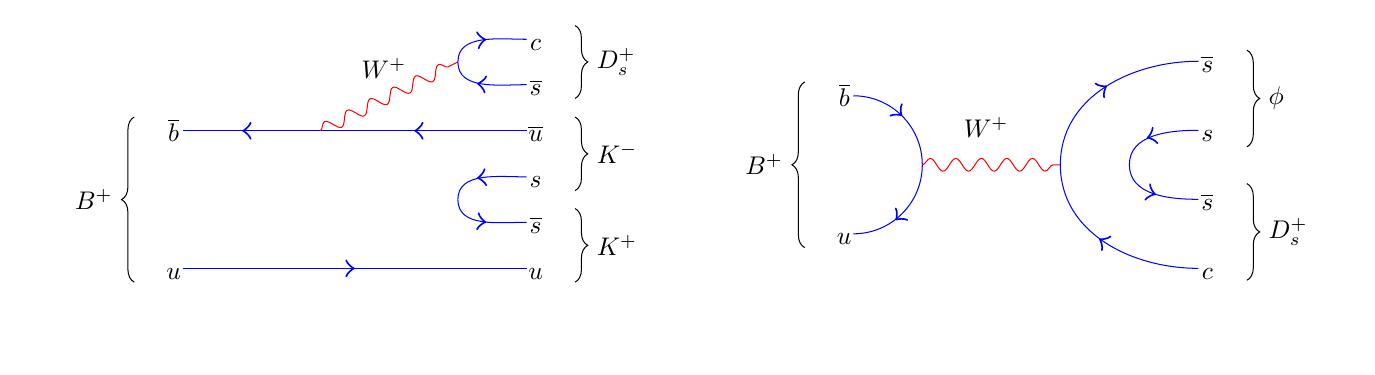}
    \caption{Dominant diagram for the (left) $\decay{\Bp}{\Dsp\Kp\Km}$ decay and (right) annihilation diagram for the $\decay{\Bp}{\Dsp\phi}$ decay in the Standard Model.}
    \label{fig:DsPhiDiagram}
\end{figure}

The decay $\decay{\Bp}{\Dsp\Kp\Km}$ is mediated by a $\decay{\bquarkbar}{\uquarkbar}$ transition shown in Fig.~\ref{fig:DsPhiDiagram} and is therefore suppressed in the Standard Model (SM) due to the small size of the Cabibbo-Kobayashi-Maskawa (CKM) matrix element \Vub. The branching fraction for this decay is currently not measured, however a similar decay, \decay{\Bp}{\Dsp \piz}, has been observed with a branching fraction of $\mathcal{B}(\decay{\Bp}{\Dsp \piz}) = (1.5 \pm 0.5) \times 10^{-5}$~\cite{Aubert:2006xy}.

In the SM, the decay $\decay{\Bp}{\Dsp\phi}$ proceeds dominantly via the annihilation diagram shown in Fig.~\ref{fig:DsPhiDiagram}. 
This suppressed topology requires the wave functions of the incoming quarks to overlap sufficiently to annihilate into a virtual \Wp boson. 
The decay is further suppressed by the small magnitude of the CKM matrix element \Vub associated with the annihilation vertex. 
In addition, unlike many rare hadronic decays including $\decay{\Bp}{\Dsp\Kp\Km}$, possible contributions from rescattering effects are expected to be small, for example contributions from intermediate states such as $\decay{\Bp}{\Dsp\omega}$~\cite{Gronau:2012gs}.
Several SM predictions have been made for the branching fraction of the $\decay{\Bp}{\Dsp\phi}$ decay~\cite{Zou:2009zza, Mohanta:2002wf, Mohanta:2007uu, Lu:2001yz}, using input from lattice calculations~\cite{fB:2013HPQCD,fB:2016ETM, fB:2016Fermi}. These predictions are in the range $(1-7)\times10^{-7}$, where the limit on the precision is dominated by hadronic uncertainties. 
However, additional diagrams contributing to this decay can arise in some extensions of the SM, such as supersymmetric models with R-parity 
violation. They could enhance the branching fraction and/or produce large \CP asymmetries~\cite{Mohanta:2002wf, Mohanta:2007uu}, which makes the $\decay{\Bp}{\Dsp\phi}$ decay a promising place to search for new physics beyond the SM.\footnote{Charge conjugation is implied throughout this paper. Furthermore, $\phi$ denotes the $\phi(1020)$ resonance.}

The \lhcb experiment reported evidence for the decay $\decay{\Bp}{\Dsp\phi}$ using $pp$ collision data corresponding to an integrated luminosity of 1\invfb taken during 2011, at a centre-of-mass energy of 7\tev~\cite{Aaij:2012zh}. A total of $6.7^{+4.5}_{-2.6}$ candidates was observed. The branching fraction was determined to be 

\begin{equation}
\mathcal{B}(\decay{\Bp}{\Dsp\phi}) = (1.87^{+1.25}_{-0.73} \pm 0.19 \pm 0.32) \times 10^{-6},
\end{equation}

\noindent where the first uncertainty is statistical, the second is systematic and the third is due to the uncertainty on the branching fraction of the decay $\decay{\Bp}{\Dsp\Dzb}$, which was used as normalisation. 
Given the large uncertainties on both the theoretical and experimental values, the previously measured value is consistent with the range of SM values given above. 
The measurements presented in this paper reanalyse the data collected in 2011, whilst adding data corresponding to an integrated luminosity of 2\invfb collected at a centre-of-mass energy 8\tev in 2012, along with 0.3\invfb from 2015 and 1.5\invfb from 2016, both at 13\tev. They supersede the previous measurement~\cite{Aaij:2012zh}.

This analysis is performed in two parts: firstly $\decay{\Bp}{\Dsp\Kp\Km}$ decays are reconstructed across the entire phase space and then a dedicated search for $\decay{\Bp}{\Dsp\phi}$ decays is performed in a narrow region of $\Kp\Km$ invariant mass around the \phiz meson.
The branching fractions are determined using the decay $\decay{\Bp}{\Dsp\Dzb}$, with $\decay{\Dzb}{\Kp\Km}$, as a normalisation channel. 
Although this \Dzb decay has a smaller branching fraction than $\decay{\Dzb}{\Kp\pim}$ (0.4\% vs. 3.9\%~\cite{PDG2016}), sharing 
the same final state between the signal and normalisation channel reduces systematic uncertainties in the ratio of detection efficiencies.

%% file: detector.tex
\section{Detector and data sample}
\label{sec:Detector}

The \lhcb detector~\cite{Alves:2008zz,LHCb-DP-2014-002} is a single-arm forward
spectrometer covering the \mbox{pseudorapidity} range $2<\eta <5$,
designed for the study of particles containing \bquark or \cquark
quarks. The detector includes a high-precision tracking system
consisting of a silicon-strip vertex detector surrounding the $pp$
interaction region, a large-area silicon-strip detector located
upstream of a dipole magnet with a bending power of about
$4{\mathrm{\,Tm}}$, and three stations of silicon-strip detectors and straw
drift tubes placed downstream of the magnet.
The tracking system provides a measurement of momentum, \ptot, of charged particles with
a relative uncertainty that varies from 0.5\% at low momentum to 1.0\% at 200\gevc.
The minimum distance of a track to a primary $pp$ interaction vertex (PV), the impact parameter (IP), 
is measured with a resolution of $(15+29/\pt)\mum$,
where \pt is the component of the momentum transverse to the beam, in\,\gevc.
Different types of charged hadrons are distinguished using information
from two ring-imaging Cherenkov detectors. 
Photons, electrons and hadrons are identified by a calorimeter system consisting of
scintillating-pad and preshower detectors, an electromagnetic
calorimeter and a hadronic calorimeter. Muons are identified by a
system composed of alternating layers of iron and multiwire
proportional chambers.
The online event selection is performed by a trigger, 
which consists of a hardware stage, based on information from the calorimeter and muon
systems, followed by a software stage, which applies a full event
reconstruction.

At the hardware trigger stage, events are required to have a muon with high \pt or a hadron, photon or electron with high transverse energy in the calorimeters. 
Two different algorithms are used in the software trigger to select candidates for this analysis.
The first uses a multivariate algorithm~\cite{BBDT} to identify the presence of a secondary vertex that has two, three or four tracks and is displaced from any PV. At least one of these charged particles must have a transverse momentum $\pt > 1.7\gevc$ and be inconsistent with originating from a PV. 
The second algorithm selects $\phiz$ candidates decaying to two charged kaons. Each kaon must have a transverse momentum $\pt > 0.8\gevc$ and be inconsistent with originating from a PV. The invariant mass of the kaon pair must be within $20\mevcc$ of the known \phiz mass~\cite{PDG2016}. This algorithm is used in both the search for $\decay{\Bp}{\Dsp\phiz}$ and $\decay{\Bp}{\Dsp\Kp\Km}$ decays.

Simulated events are used to determine the relative efficiencies of the signal and normalisation channels. 
The samples are generated for each of the running periods. In these simulations, $pp$ collisions are generated using \pythia~\cite{Sjostrand:2007gs,Sjostrand:2006za} with a specific \lhcb configuration~\cite{LHCb-PROC-2010-056}.  Decays of hadronic particles are described by \evtgen~\cite{Lange:2001uf}, in which final-state radiation is generated using \photos~\cite{Golonka:2005pn}. The interaction of the generated particles with the detector, and its response, are implemented using the \geant toolkit~\cite{Allison:2006ve, *Agostinelli:2002hh} as described in Ref.~\cite{LHCb-PROC-2011-006}.

%% file: selection.tex
\section{Candidate selection}
\label{sec:Selection}


Candidate $\decay{\Bp}{\Dsp\phiz}$ and $\decay{\Bp}{\Dsp\Kp\Km}$ decays are selected using similar requirements. The \phiz mesons in $\decay{\Bp}{\Dsp\phiz}$ candidates are reconstructed with $\decay{\phi}{\Kp\Km}$.
Both modes are reconstructed using the $\decay{\Dsp}{\Kp\Km\pip}$ decay, whilst $\decay{\Bp}{\Dsp\phiz}$ candidates are additionally reconstructed with the decays $\decay{\Dsp}{\Kp\pim\pip}$ and $\decay{\Dsp}{\pip\pim\pip}$ to increase the sensitivity of the search. 
The $\Dsp$ ($\phi$) candidates are required to have an invariant mass within $25\mevcc$ ($40\mevcc$) of the known $\Dsp$ ($\phiz$) mass~\cite{PDG2016}. In the search for $\decay{\Bp}{\Dsp\Kp\Km}$ decays, the veto $|m(\Kp\Km)-m(\Dz)| > 25\mevcc$ is applied to explicitly remove the normalisation channel from the signal mode.

The \Bp meson candidates are formed from well reconstructed tracks with $\chisqip > 4.0 $, where \chisqip is defined as the difference in the vertex-fit \chisq of the best PV reconstructed with and without the particle being considered. The best PV is the PV that has the smallest \chisqip value. For kaons from the \phiz or \Bp decay the momentum requirement is $p > 2 \gevc $.  At least one track of each \Bp meson candidate must have $\pt > 0.5 \gevc $ and $p > 5\gevc$.

Loose requirements are made on particle identification (PID) to reduce background from other \bquark-hadron decays with misidentified hadrons. For the signal, the overall efficiency of the PID requirements varies from 80\% to 90\%, depending on the \Dsp mode.
Background from decays of \Bp mesons to the same final state that did not proceed via a \Dsp meson (referred to as charmless decays) are suppressed by applying a requirement on the significance of the \Bp and \Dsp vertex separation, $\chi^{2}_{\text{FD}}$. 

The residual yields of charmless decays are estimated by determining the \Bp yield in candidates that are in the invariant mass range $25 < |m(h^{+}h'^{-}\pip) - m(\Dsp)| < 50\mevcc $, where $m(h^{+}h'^{-}\pip)$ is the \Dsp candidate mass and $h,h'=K,\pi$. This background estimation is performed separately for the $\decay{\Bp}{\Dsp\phiz}$ and $\decay{\Bp}{\Dsp\Kp\Km}$ searches. For the $\decay{\Bp}{\Dsp\Dzb}$ normalisation channel, a two-dimensional optimisation is performed to calculate the contribution from decays without a \Dsp meson, \Dzb meson or both. The optimal selection requirements are chosen such that the maximum signal efficiency is achieved for a residual charmless contribution of 2\% of the normalisation yield.

For the decay $\decay{\Dsp}{\Kp\Km\pip}$, candidates are rejected if they are consistent with $\decay{\Dp}{\Km\pip\pip}$ or $\decay{\Lc}{\proton\Km\pip}$ decays, where either a pion or a proton has been misidentified as a kaon. The candidates are reconstructed using the alternative mass hypothesis and, for those falling within $25\mevcc$ of the \Dp or \Lc mass, particle identification requirements are tightened on the misidentified track.

Another set of vetoes rejects decays where the tracks forming the \Dsp candidate originate from an excited charged charm meson decay, for example $\decay{\Dstarp}{(\decay{\Dz}{h^{+}h'^{-}}) \pip}$. By requiring $\Delta m = m(h^{+}h'^{-}\pip)-m(h^{+}h'^{-}) > 150 \mevcc$ decays of this type are efficiently removed. Other specific backgrounds are removed by mass vetoes. These vetoes remove $\decay{\Bs}{\phiz\phiz}$ decays in which one of the \phiz mesons is combined with an unrelated pion to form the \Dsp candidate. Any candidates within $50\mevcc$ of the known \Bs mass~\cite{PDG2016} in the four-body invariant mass $m(\Kp\Km\Kp\Km)$ are removed to ensure a smooth combinatorial background distribution. 

In addition, a veto is applied to the invariant mass of the kaons from the \phiz meson or \Bp candidate combined with any pion from the \Dsp candidate, removing candidates within $25\mevcc$ of the known \Dsp mass. This removes decays that include incorrectly reconstructed $\decay{\Dsp}{\phi\pip}$ or $\decay{\Dsp}{\Kp\Km\pip}$ decays, where the $\phiz$ or $\Kp\Km$ pair are incorrectly assigned to have originated from the \Bp meson rather than the \Dsp meson. For example, this incorrect assignment could lead to $\decay{\Bp}{(\decay{\Dsp}{\phiz\pip})\Kp\Km}$ decays being reconstructed as $\decay{\Bp}{(\decay{\Dsp}{\Kp\Km\pip}) \phiz}$ decays. 
The \Bp (\Dsp) candidates are required to have $\chisqip < 10$ ($\chisqip > 10$), to ensure they are consistent (inconsistent) with being produced at the best PV.

Multivariate analyses (MVA) are used to separate genuine \phiz and \Dsp candidates from random combinations of tracks~\cite{LHCb-PAPER-2012-050}. The \phiz and \Dsp MVAs use data samples of $\decay{\Bs}{\jpsi \phiz}$ and $\decay{\Bsb}{\Dsp \pim}$ decays, respectively, where the background is statistically subtracted using the \sPlot method~\cite{Pivk:2004ty}. The training uses the \phiz or \Dsp sidebands as a background sample. A total of eight MVAs are trained to target the decays $\decay{\phi}{\Kp \Km}$, $\decay{\Dsp}{\Kp\Km\pip}$, $\decay{\Dsp}{\Kp\pim\pip}$ and $\decay{\Dsp}{\pip\pim\pip}$, separately in the Run 1 (2011 and 2012) and Run 2 (2015 and 2016) data. A preselection including the trigger, vetoes and PID requirements previously discussed is applied to the training samples, ensuring they are representative of the target signal decays. The samples are split into two subsamples in a random but reproducible way. One is used to train the corresponding MVA, the other to test its response. 

The MVA method used in this analysis is a gradient Boosted Decision Tree (BDTG)~\cite{Breiman}. The selection criteria for each of the BDTG classifiers are determined by optimising the figure of merit $\epsilon_{s}/ (\frac{a}{2} + \sqrt{N_{\text{BKG}}})$~\cite{Punzi:2003bu}, with $a=5$, where $\epsilon_{s}$ is the signal efficiency and $N_{\text{BKG}}$ is the number of background candidates determined from fits to data, calculated in the signal region.

The efficiencies of the MVAs are obtained from the test samples of $\decay{\Bs}{\jpsi\phiz}$ and $\decay{\Bsb}{\Dsp\pim}$ decays. Additionally, a sample of $\decay{\Bp}{\Dzb\pip}$ decays is used to calculate the efficiency of $\decay{\Dzb}{\Kp\Km}$ decays in the normalisation channel. The efficiency calculation takes into account the kinematic differences between the training and signal samples, as well as any possible correlations between the \Dsp and \phiz kinematics, by using input from simulation samples. Any further correlations between the \phiz and \Dsp MVA efficiencies are found to be negligible. In the search for $\decay{\Bp}{\Dsp\Kp\Km}$ decays, calibration samples are used to correct for the imperfect modelling of the PID in simulation. These corrected samples are then used to obtain the variations in the MVA efficiencies as a function of the phase-space position, in particular of the $m(\Kp\Km)$ invariant mass.

The invariant mass of the \Bp meson candidates is determined from fits in which the \Dsp candidate mass (and \Dzb candidate mass for the normalisation channel) is constrained to the known value~\cite{Hulsbergen:2005pu}. Additionally, the momentum vector of the \Bp meson is constrained to be parallel to the vector connecting the PV and the \Bp meson decay vertex.

%% file: fit.tex
\section{Invariant mass fits}
\label{sec:Fit}

The branching fractions of the $\decay{\Bp}{\Dsp\phiz}$ and $\decay{\Bp}{\Dsp\Kp\Km}$ decays are determined from unbinned maximum likelihood fits to the invariant mass of the \Bp candidates. However, separate fit strategies are used for the $\decay{\Bp}{\Dsp\phiz}$ and $\decay{\Bp}{\Dsp\Kp\Km}$ searches.

The search for $\decay{\Bp}{\Dsp\Kp\Km}$ involves two independent fits for the signal and normalisation channels. The $\decay{\Bp}{\Dsp\Kp\Km}$ yield is corrected on a per-candidate basis to account for the phase-space dependence of the signal efficiencies in this three-body decay. 

\begin{figure}[t]
    \centering
    \begin{subfigure}[t]{1.0\textwidth}
    \centering 
        \includegraphics[width=1.0\textwidth]{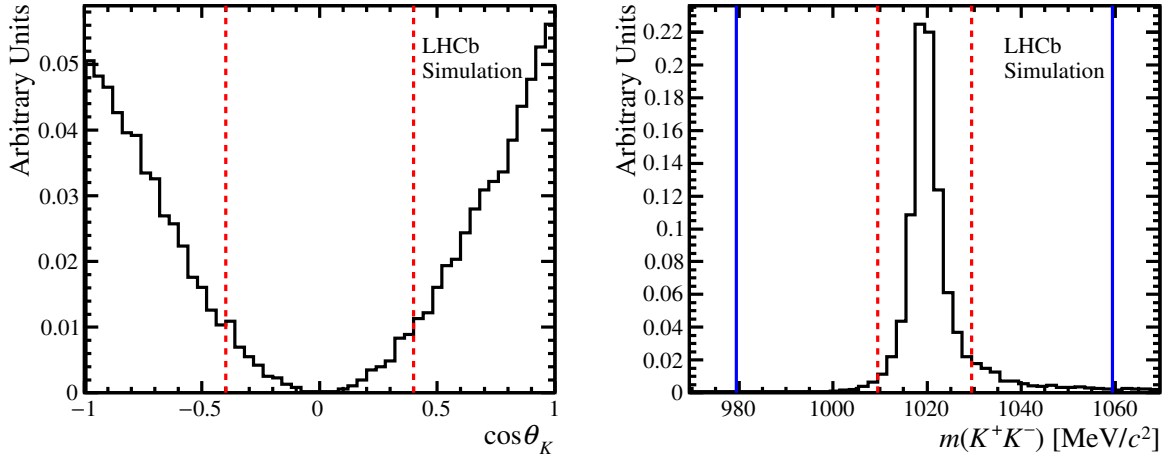}
    \end{subfigure}  
    \caption{Distributions of (left) $\cos{\theta_{K}}$ and (right) $m(\Kp\Km)$ in $\decay{\Bp}{\Dsp\phiz}$ decays, as determined from simulated events. The vertical lines represent the limits of the two categories used for each variable. In the $m(\Kp\Km)$ distribution, the area within the dashed red lines represents the \phiz signal region, and the two areas between the dashed red and blue lines represent the \phiz sideband region. The $\decay{\Bp}{\Dsp \phiz}$ signal decays are seen to primarily contribute to the $\phi$ signal region and the $|\cos{\theta_{K}}| > 0.4$ category. } 
    \label{fig:DsPhiMC}   
\end{figure}

In contrast, the $\decay{\Bp}{\Dsp\phiz}$ candidates are treated as quasi-two-body decays in which all signal candidates are corrected with the same efficiency. The $\decay{\Bp}{\Dsp\phiz}$ signal and normalisation channels are fitted simultaneously in different categories, as are the three \Dsp decay modes,  with the $\decay{\Dsp}{\Kp\Km\pip}$ mode split further into $\decay{\Dsp}{\phi \pip}$ and non-\phiz submodes.
This exploits the high purity of the $\decay{\Dsp}{\phi \pip}$ decay.
As the $\decay{\Bp}{\Dsp\phiz}$ decay involves the decay of a pseudoscalar particle to a pseudoscalar and vector particle, the \phiz vector meson ($J^{P} = 1^{-}$) must be produced longitudinally polarised. For a longitudinally polarised \phiz meson decaying to $\Kp\Km$, the distribution of the angle $\theta_{K}$, defined as the angle that the kaon meson forms with the \B momentum in the \phiz rest frame, is proportional to $\cos^{2}{\theta_{K}}$. The distribution of $\cos{\theta_{K}}$ for $\decay{\Bp}{\Dsp\phiz}$ as determined from simulated events is shown in Fig.~\ref{fig:DsPhiMC}.
In the simultaneous fit for $\decay{\Bp}{\Dsp \phiz}$ candidates the candidates are split into two helicity categories: $|\cos{\theta_{K}}|>0.4$ and $|\cos{\theta_{K}}|<0.4$. In simulated events, 93\% of $\decay{\Bp}{\Dsp\phiz}$ decays are found in the first category, whereas for the normalisation decay and background modes, as the distributions in $\cos{\theta_{K}}$ are approximately flat, only 60\% of candidates fall into this category. Additionally, the fit further assigns candidates into two $m(\Kp\Km)$ invariant mass categories, $|m(\Kp\Km) - m_{\phi}|<10 \mevcc$ and  $10 <|m(\Kp\Km) - m_{\phi}|< 40 \mevcc$ (Fig.~\ref{fig:DsPhiMC}), to help constrain the contribution from the different backgrounds in the signal region. Background modes involving two kaons that did not originate from a \phiz meson (for example $\decay{\Bsb}{D_{s}^{(*)+} K^{-} \Kstarz}$) have different fractions in these two categories, helping to distinguish them from those decays with a real $\phi$ meson. The fractions of $\decay{\Bp}{\Dsp\phiz}$ candidates in each of the categories, as determined from simulated events, are listed in Table~\ref{tab:signal_ratios}.

\begin{table}[t]
  \caption{Fractions of $\decay{\Bp}{\Dsp\phiz}$ candidates expected in the helicity and $m(\Kp\Km)$ invariant mass categories of the simultaneous fit. }
  \label{tab:signal_ratios}

  \begin{center}\begin{tabular}{c|cc}
    \hline
    \multirow{2}{*}{$| m(\Kp\Km) - m_{\phi} |$ (\mevcc)}   & \multicolumn{2}{c}{Helicity Category} \\ 
     & $|\cos{\theta_{K}} |> 0.4$          & $|\cos{\theta_{K}} |< 0.4$\\ 
    \hline
    $< 10$                           & 82\%           & 6\%                       \\
    (10, 40)                         & 11\%           & 1\%                       \\
    \hline
  \end{tabular}\end{center}
\end{table}

\begin{table}[t]

  \caption{Fractions of $\decay{\Bp}{\Dsp\Kp\Km}$ candidates assumed to contribute to each helicity and $m(\Kp\Km)$ invariant mass categories of the simultaneous fit. The uncertainties shown are calculated from the range of fractions obtained by assuming different contributing resonances, as detailed in Sec.~\ref{sec:fitcomponents}.}
  \label{tab:DsKK_ratios}

  \begin{center}\begin{tabular}{c|cc}
    \hline
    \multirow{2}{*}{$| m(\Kp\Km) - m_{\phi} |$ (\mevcc)}   & \multicolumn{2}{c}{Helicity Category} \\ 
     & $|\cos{\theta_{K}} |> 0.4$          & $|\cos{\theta_{K}} |< 0.4$\\ 
    \hline
    $< 10$                           & $(15\pm2)$\%           & $(10\pm1)$\%                       \\
    (10, 40)                         & $(45\pm2)$\%           & $(30\pm1)$\%                       \\
    \hline
  \end{tabular}\end{center}
\end{table}

\subsection{Signal and normalisation probability density functions}
\label{sec:fitcomponents}
The normalisation and signal components in the $\decay{\Bp}{\Dsp\Dzb}$ and $\decay{\Bp}{\Dsp\Kp\Km}$ or $\decay{\Bp}{\Dsp\phi}$ invariant mass distributions are each modelled using the sum of two Crystal Ball (CB)~\cite{Skwarnicki:1986xj} probability density functions (PDFs) with tails at lower invariant mass. The tail parameters, the ratio of the two CB widths, and the relative fraction of each CB function are determined from simulated events. 
The resolution parameter of the narrow CB component in each \Dsp decay mode category is a free parameter in the fit, but the ratios of signal and normalisation widths are fixed to values determined from simulated events. For the normalisation mode, the fraction of $\decay{\Bp}{\Dsp\Dzb}$ candidates in the two helicity bins is a free parameter in the fit, whereas for the signal the fraction in each helicity and $m(\Kp\Km)$ invariant mass category of the fit is fixed to that determined from simulated events, as reported in Table~\ref{tab:signal_ratios}.

The search for $\decay{\Bp}{\Dsp \phiz}$ decays includes a component for $\decay{\Bp}{\Dsp\Kp\Km}$ decays that did not proceed via a \phiz meson. The fraction of $\decay{\Bp}{\Dsp\Kp\Km}$ decays expected in each helicity angle and $m(\Kp\Km)$ mass category, shown in Table~\ref{tab:DsKK_ratios}, are calculated from the average of different $\Kp\Km$ resonances that could contribute to $\decay{\Bp}{\Dsp\Kp\Km}$ decays. These resonances include possible contributions from the $f_{0}(980)$ and $a_{0}(980)$ resonances. 
The resulting fractions are sufficiently different from those for the $\decay{\Bp}{\Dsp \phiz}$ signal such that the two contributions can be distinguished. The range of fractions obtained by considering the different resonances are included as uncertainties in Table~\ref{tab:DsKK_ratios}. A systematic uncertainty is assigned to account for the fixed fractions assumed in the fit. No attempt is made to separate any of the contributing resonances in the search for $\decay{\Bp}{\Dsp\Kp\Km}$ candidates. 

\subsection{Background PDFs}

\begin{figure}[t]
    \centering
    \begin{subfigure}[t]{0.69\textwidth}
    \centering 
        \includegraphics[width=1.0\textwidth]{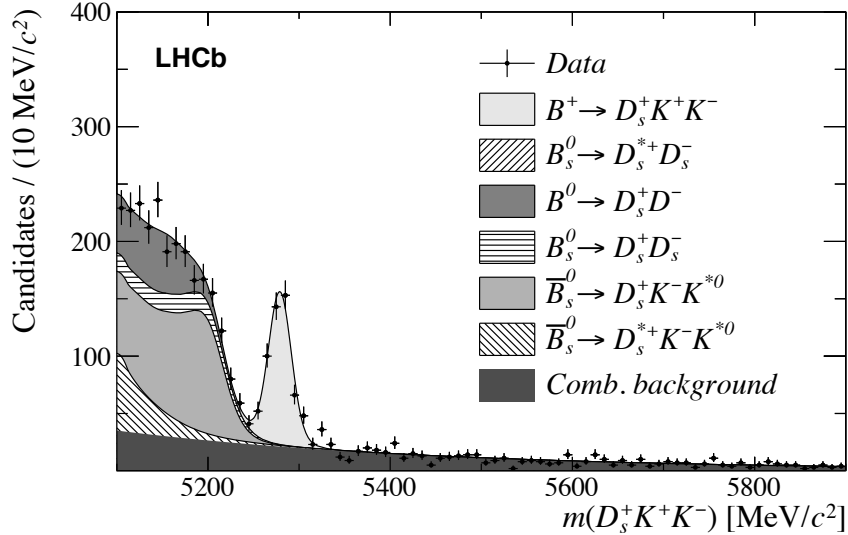} 
    \end{subfigure}  
    \caption{Mass distribution of $\decay{\Bp}{\Dsp\Kp\Km}$ candidates. The result of the fit to the data using the model described in Sec.~\ref{sec:fitcomponents} is overlaid, with the PDF components given in the legend.  } 
    \label{fig:DsKKSignalfit}   
\end{figure}

\begin{figure}[t]
    \centering
    \begin{subfigure}[t]{0.49\textwidth}
    \centering 
        \includegraphics[trim={6cm 0 4cm 0},width=1.0\textwidth]{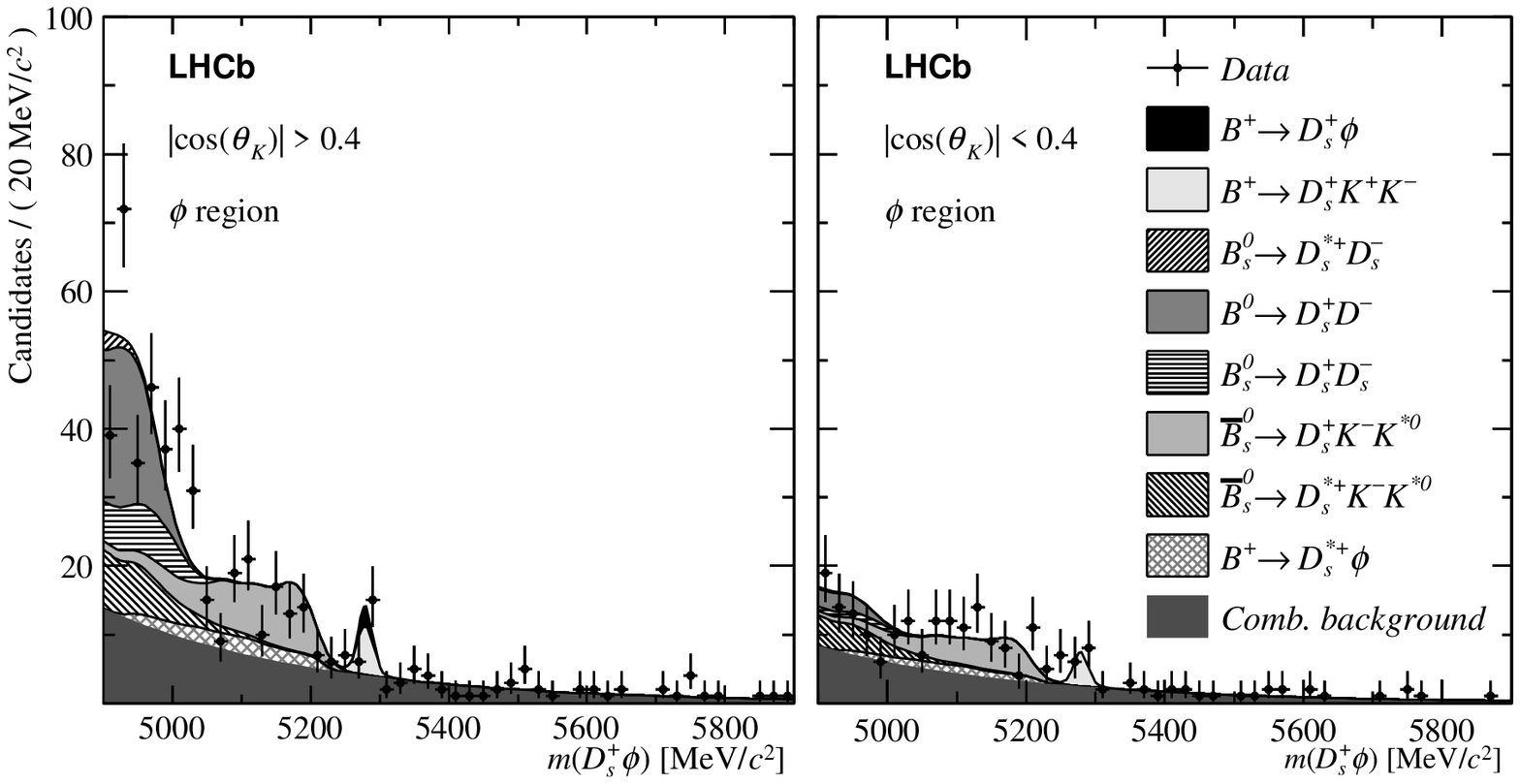}
    \end{subfigure}\\
    \begin{subfigure}[t]{0.49\textwidth}    
        \centering 
        \includegraphics[trim={6cm 0 4cm 0},width=1.0\textwidth]{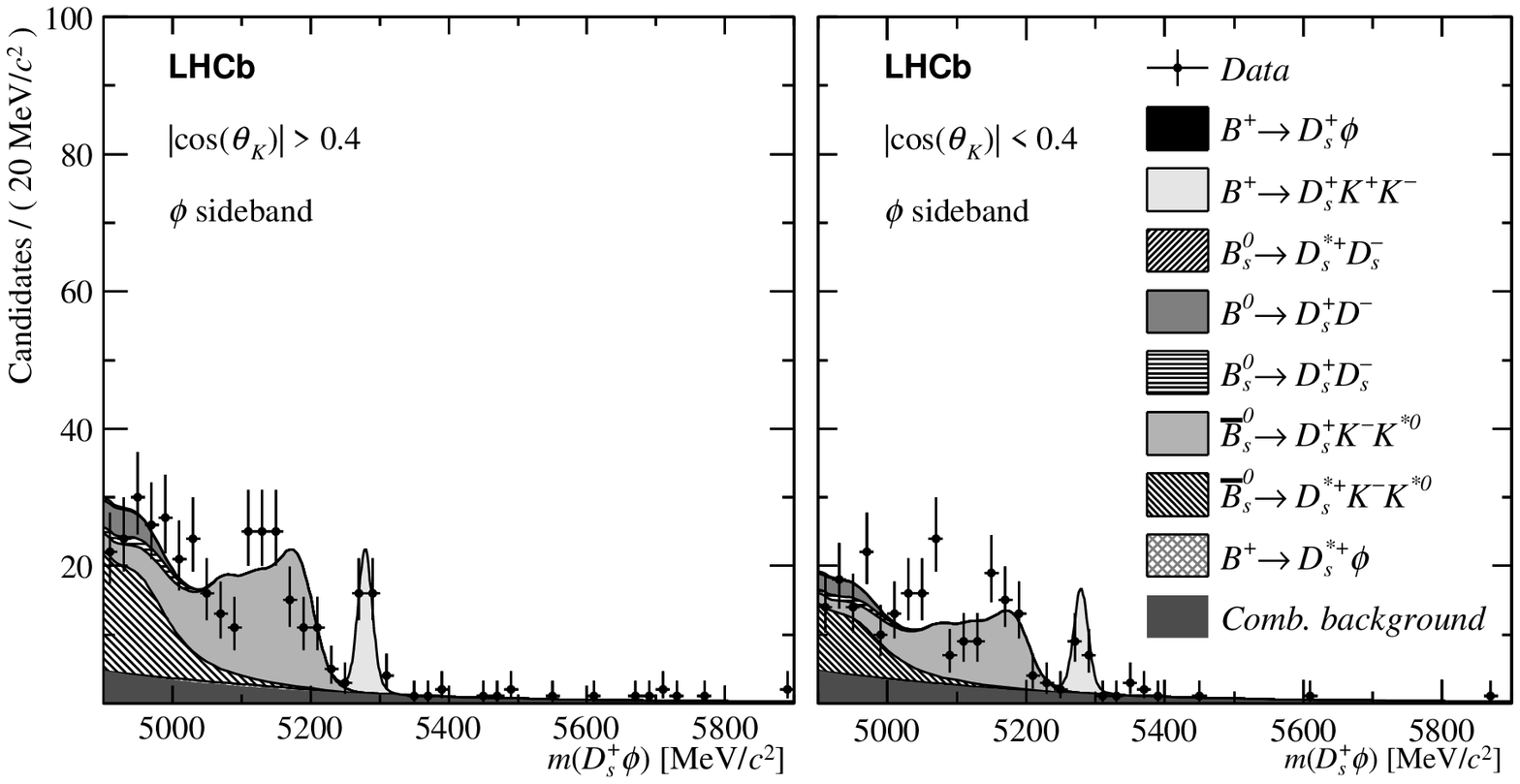}
    \end{subfigure}\\   
    \caption{Mass distribution of $\decay{\Bp}{\Dsp \phiz}$ candidates in (top) the \phiz mass region, and (bottom) the \phiz mass sideband. The plots on the left are in the helicity bin $|\cos{\theta_{K}}|>0.4$ and the right are in $|\cos{\theta_{K}}|<0.4$. The result of the fit to the data using the model described in Sec.~\ref{sec:fitcomponents} is overlaid, with the PDF components given in the legend. The $\decay{\Bp}{\Dsp \phiz}$ decays (black) are expected to primarily contribute to the $\phi$ region with $|\cos{\theta_{K}}| > 0.4$.} 
    \label{fig:Signalfit}   
\end{figure}

A number of background components are included in the fit model. The dominant source of background under the signal is due to combinations of unrelated tracks. 
An exponential function is used to parametrise this component. The same slope parameter is used in the simultaneous fit to the signal and normalisation modes.
Partially reconstructed $\decay{\Bp}{\Dssp\Dzb}$ and $\decay{\Bp}{\Dsp\Dstarzb}$ decays are concentrated in the lower part of the $\Dsp\Dzb$ spectrum. 
They are parametrised using analytical shapes that account for the nonreconstructed neutral pion or photon from the excited \D-meson decays. These shapes are constructed from Gaussian distributions convolved with second-order polynomials, and are analogous to those used in similar analyses~\cite{LHCb-PAPER-2017-021}. An additional component is used to model $\decay{\Bp} {\Dss\Dstarzb}$ decays where one particle from each of the excited \D mesons is missed.
Partially reconstructed $\decay{\Bp}{\Dssp\phiz}$ decays can contribute to the lower part of the $\Dsp \phi$ spectrum. These, similarly, are fitted with analytical shapes that account for the missing neutral particle from the $\Dssp$ decay, as well as the different helicity states for the decay of a pseudoscalar meson to two vector particles.
They are parametrised in an analogous way to similar analyses~\cite{LHCb-PAPER-2016-006}. This background component is only included in the search for $\decay{\Bp}{\Dsp \phiz}$ decays.
The modes $\decay{\Bsb}{\Dsp\Km\Kstarz}$ and $\decay{\Bsb}{\Dssp\Km\Kstarz}$ form a background to $\decay{\Bp}{\Dsp \phiz}$ decays when a low-momentum pion from the \Kstarz decay is not reconstructed. Additionally, a neutral pion or photon can be missed from the excited \Dsp meson decay in the case of $\decay{\Bsb}{\Dssp\Km\Kstarz}$. The PDFs are determined from simulated events. The expected fractions in each category of the $\decay{\Bp}{\Dsp \phiz}$ fit are fixed using simulated events. 
The decays $\decay{\Bs}{\Dsp\Dsm}$, $\decay{\Bs}{\Dssp\Dsm}$ and $\decay{\Bd}{\Dsp\Dm}$ can form a background when a pion is not reconstructed from a \Dsp or \Dp decaying to $\Kp \Km \pip$. The PDFs are also determined from simulated events, with the fractions in each $\decay{\Bp}{\Dsp \phiz}$ fit category fixed.
The result of the fit to $\decay{\Bp}{\Dsp\Kp\Km}$ candidates, including all the relevant background components is shown in Fig.~\ref{fig:DsKKSignalfit}.
The result of the simultaneous fit to $\decay{\Bp}{\Dsp \phiz}$ candidates in the different helicity angle and $m(\Kp\Km)$ mass categories is shown in Fig.~\ref{fig:Signalfit}. The three contributing \Dsp meson decay modes are merged.

%% file: systematics.tex
\section{Systematic uncertainties}
\label{sec:systematics}

A number of different sources of systematic uncertainty are considered. The contribution from each source is detailed in Table~\ref{tab:sys}.

\begin{table}[t]
  \caption{Systematic uncertainties contributing to the measurements of $\mathcal{B}(\decay{\Bp}{\Dsp\phiz})$ and $\mathcal{B}(\decay{\Bp}{\Dsp\Kp\Km})$. The systematic uncertainty from the normalisation branching fraction is also included.  }
\begin{center}\begin{tabular}{l c c}
    \hline
    \multirow{2}{*}{Source of uncertainty} &$\mathcal{B}(\decay{\Bp}{\Dsp\phiz})$  &$\mathcal{B}(\decay{\Bp}{\Dsp\Kp\Km})$ \\ 
                                           &($\times 10^{-7}$)                     &($\times 10^{-6}$)                     \\ 
    \hline
    Relative efficiencies                     & 0.08   & 0.59 \\
    Signal and normalisation PDFs             & 0.04   & 0.04 \\
    Background PDFs                           & 0.69   & 0.02 \\
    Charmless contribution                    & 0.02   & 0.05 \\
    $\decay{\Bp}{\Dsp\Kp\Km}$ model   & 0.38  & -- \\
    \hline
    Normalisation                             & 0.12  & 0.72 \\
    \hline
  \end{tabular}\end{center}
\label{tab:sys}
\end{table}

\begin{description}

\item \textbf{Relative efficiencies:} The calculation of the branching fractions requires a correction to the ratio of signal and normalisation yields to account for the difference in the selection efficiency of the two modes. All relative selection efficiencies except the PID and MVA efficiencies are determined from simulated events and the effect of having a limited simulation sample size is included as a systematic uncertainty.
The relative efficiency for the PID and MVA requirements are determined from data control modes, including the samples of \decay{\Bs}{\jpsi \phiz} and \decay{\Bsb}{\Dsp \pim} decays used to test the MVA responses. Systematic uncertainties are assigned to account for the limited sizes of the control mode samples, kinematic differences between the control modes and the signal modes and differences between the data and simulation distributions that might affect the relative efficiency.

\item \textbf{Signal and normalisation PDFs:} Some parameters in the signal and normalisation PDFs are fixed to values obtained from simulation. These include the tail parameters, relative widths, and fractional amounts of the two CB functions that make up the PDFs. The values obtained from simulation have associated uncertainties arising from the limited simulation sample sizes. The nominal fits are repeated with the fixed parameters modified to values sampled from Gaussian distributions, with a width given by the parameter uncertainties. All parameters are changed simultaneously. For the fit to $\decay{\Bp}{\Dsp\phiz}$ candidates, the fractions of events expected in each category of the fit are also included in the procedure. The resulting variation is assigned as the systematic uncertainty.  

\item \textbf{Background PDFs:} Some of the PDFs for the background modes are taken directly from simulated events using one-dimensional kernel estimations~\cite{Cranmer:2000du}. In the nominal fit, these are smeared to account for the differences in the mass resolution between data and simulation. To account for any systematic uncertainty arising from the choice of resolution difference, the fit is repeated, randomly varying the smearing resolution each time. The resulting variation in the branching fraction is assigned as a systematic uncertainty. Additionally, each partially reconstructed background PDF has fixed fractions in the different categories of the signal fit. To determine the effect on the branching fraction, these fractions are repeatedly sampled from Gaussian distributions with widths given by the statistical uncertainty on the fractions. For the combinatorial background shape, the choice of parametrisation is varied and the effect included in the systematic uncertainty. 

\item \textbf{Charmless contribution:} Residual charmless and single-charm backgrounds are expected to remain in the final selection. These contributions are neglected in the calculation of the branching fractions. However, the shift in the branching fraction caused by numerically including the charmless yields is assigned as a systematic uncertainty.

\item \textbf{\boldmath{$\decay{\Bp}{\Dsp\Kp\Km}$} model assumption:} The fit to $\decay{\Bp}{\Dsp\phiz}$ candidates includes a shape for $\decay{\Bp}{\Dsp\Kp\Km}$ decays that do not proceed via a \phiz meson. In order to distinguish this component from the signal, the different fractions of candidates in the four fit categories are exploited. This requires making assumptions as to which resonances contribute to the full $\decay{\Bp}{\Dsp\Kp\Km}$ decay model. The shape is assumed to be dominated by $f_{0}(980)$ and $a_{0}(980)$ resonances.
Estimates of the uncertainties on the fractions are determined by considering the range in each fraction for the models considered. 
The variation in the branching fraction that results from varying these fractions within the uncertainties is assigned as the systematic uncertainty. 
\end{description}

%% file: results.tex
\section{Results}
\label{sec:Results}

\subsection{Search for \boldmath{$\decay{\Bp}{\Dsp\Kp\Km}$} candidates}

\begin{figure}[t]
    \centering
    \begin{subfigure}[t]{0.48\textwidth}
        \includegraphics[width=1.0\textwidth]{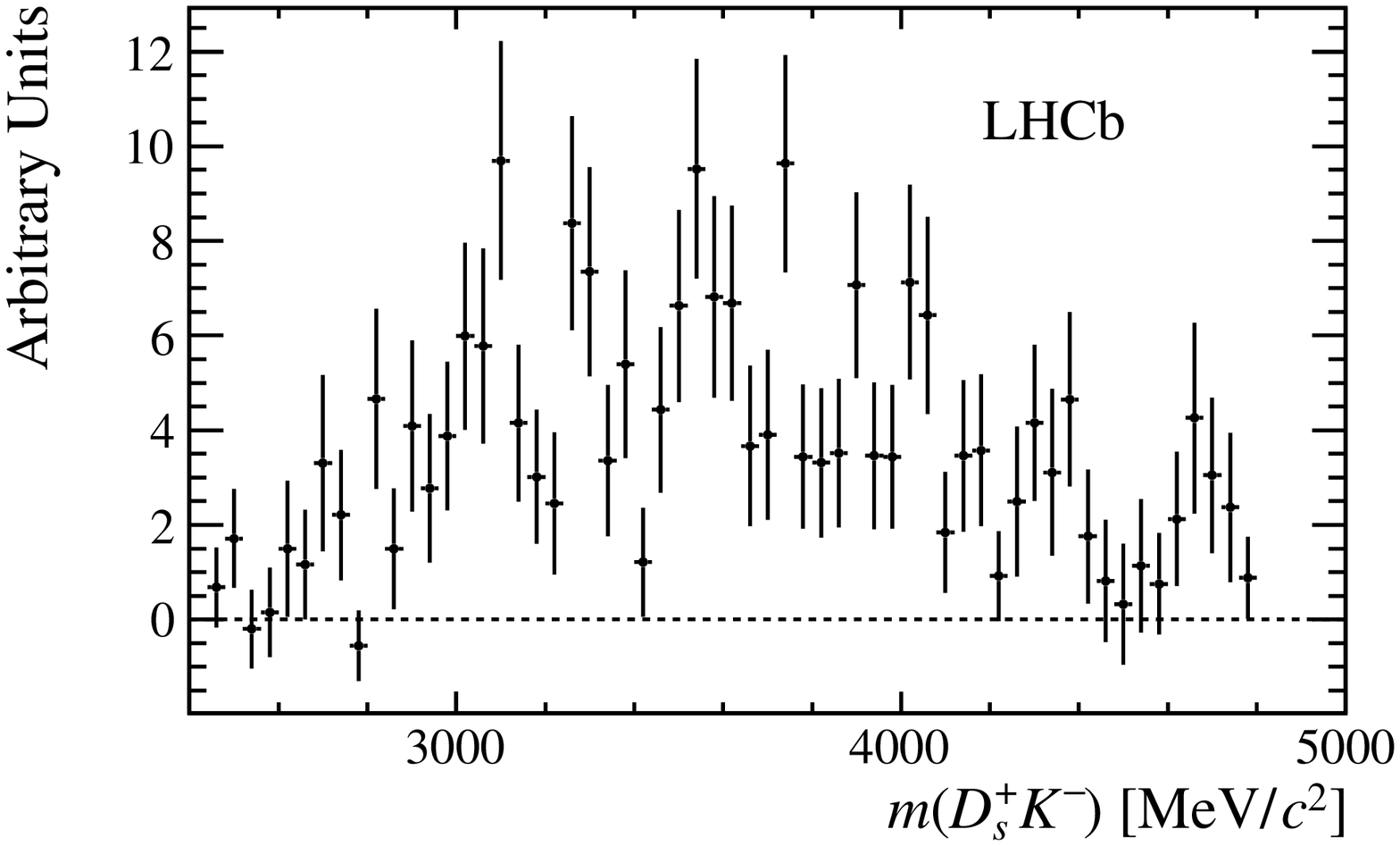}
    \end{subfigure}
    \begin{subfigure}[t]{0.48\textwidth}
        \includegraphics[width=1.0\textwidth]{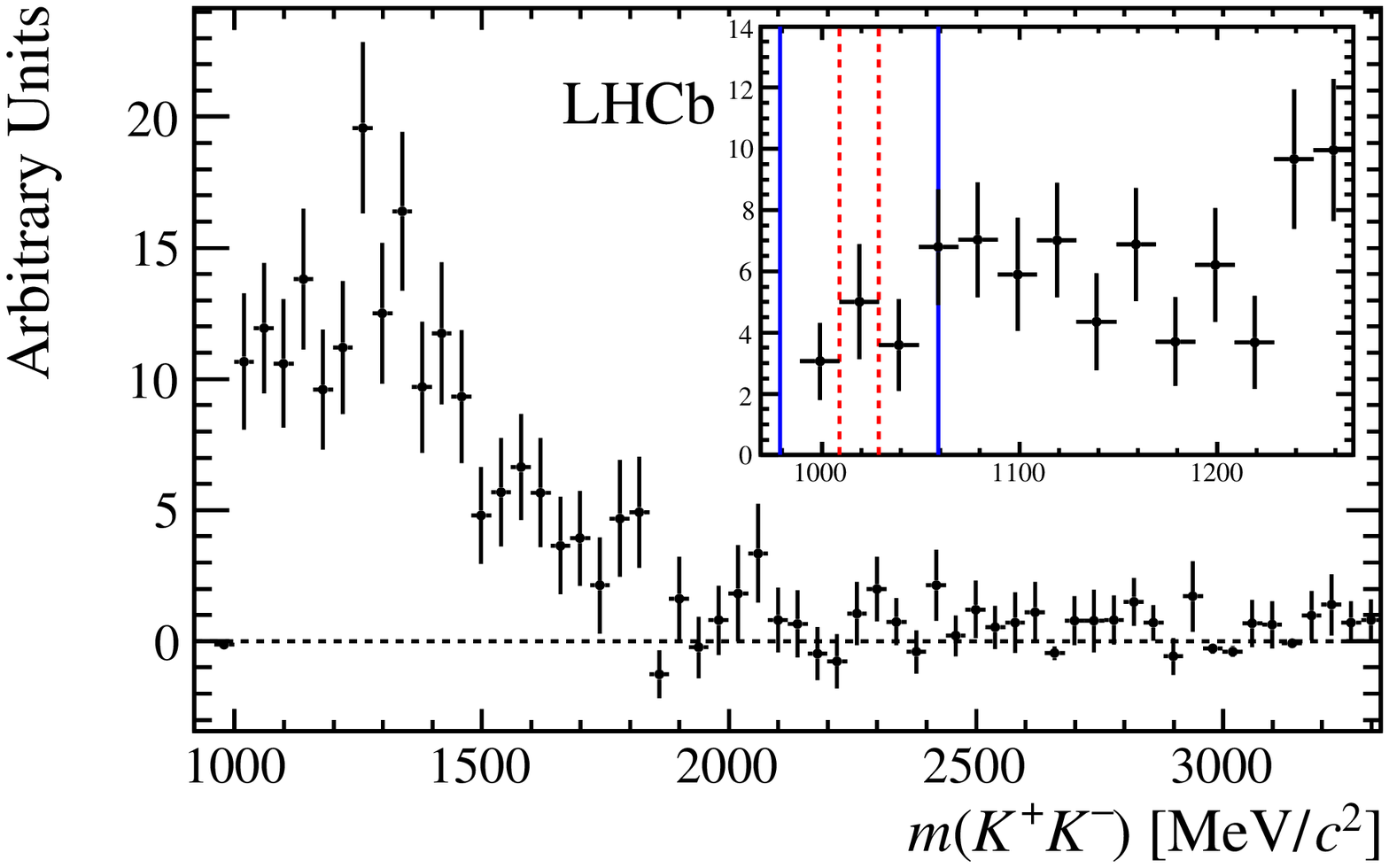}
    \end{subfigure}
    \caption{Projections of the background-subtracted two-body invariant masses (left) $m(\Dsp\Km)$ and (right) $m(\Kp\Km)$ for $\decay{\Bp}{\Dsp\Kp\Km} $ decays. These plots are additionally weighted by a factor $1/\epsilon^{\text{ratio}}_{i}$ to correct for the efficiency variation across the phase space. An expansion of the \phiz region of $m(\Kp\Km)$ is inset where the same \phiz signal region and \phiz sideband region have been represented as in Fig.~\ref{fig:DsPhiMC}.}
    \label{fig:DsKK_Projections}
\end{figure}

The fit to $\decay{\Bp}{\Dsp\Kp\Km}$ candidates finds a total yield of $N(\decay{\Bp}{\Dsp\Kp\Km}) = 443 \pm 29 $ candidates. 
This constitutes the first observation of this decay mode.
The branching fraction is calculated as
\begin{equation}
\mathcal{B}(\decay{\Bp}{\Dsp\Kp\Km}) = \frac{ N_{\text{corr}}(\decay{\Bp}{\Dsp\Kp\Km}) }{ N(\decay{\Bp}{\Dsp\Dzb}) } \times \mathcal{B}(\decay{\Bp}{\Dsp\Dzb}) \times \mathcal{B}(\decay{\Dzb}{\Kp\Km})
\label{eq:DsKKBranchingfraction}
\end{equation}
\noindent where $N(\decay{\Bp}{\Dsp\Dzb})$ is the yield of normalisation decays, and $N_{\text{corr}}(\decay{\Bp}{\Dsp\Kp\Km})$ is defined to be
\begin{equation}
N_{\text{corr}}(\decay{\Bp}{\Dsp\Kp\Km}) =  \sum\limits_{i} \frac{W_{i}}{\epsilon^{\text{ratio}}_{i}},
\end{equation}
\noindent where $W_{i}$ is the per-candidate weight, as determined by the \sPlot technique for candidate $i$; and $\epsilon^{\text{ratio}}_{i}$ represents the relative efficiency of the signal and normalisation modes $\epsilon_{i}(\decay{\Bp}{\Dsp\Kp\Km})/\epsilon(\decay{\Bp}{\Dsp\Dzb})$ in the relevant bin of the $\decay{\Bp}{\Dsp\Kp\Km}$ Dalitz plot. The corrected yield ratio can be expressed as the ratio of signal and normalisation branching fractions using Eq.~\ref{eq:DsKKBranchingfraction}. The value is measured to be 
\begin{equation*}
\frac{ N_{\text{corr}}(\decay{\Bp}{\Dsp\Kp\Km}) }{ N(\decay{\Bp}{\Dsp\Dzb}) } = \frac{\mathcal{B}(\decay{\Bp}{\Dsp\Kp\Km})}{\mathcal{B}(\decay{\Bp}{\Dsp\Dzb})\mathcal{B}(\decay{\Dzb}{\Kp\Km})}  = 0.197 \pm 0.015 \pm 0.017, 
\end{equation*}
where the first uncertainty is statistical, and the second is systematic.
The branching fraction for $\decay{\Bp}{\Dsp\Kp\Km}$ decays is determined to be 

\begin{equation*}
\mathcal{B}(\decay{\Bp}{\Dsp\Kp\Km}) = (7.1 \pm 0.5 \pm 0.6 \pm 0.7) \times 10^{-6},
\end{equation*}
where the first uncertainty is statistical, the second is systematic and the third from the branching fractions of $\decay{\Dzb}{\Kp\Km}$ and of the normalisation mode $\decay{\Bp}{\Dsp\Dzb}$. 
The values used for the branching fractions are $\mathcal{B}(\decay{\Dz}{\Kp\Km}) = (4.01 \pm 0.07)\times10^{-3}$ and $\mathcal{B}(\decay{\Bp}{\Dsp\Dzb}) = (9.0 \pm 0.9)\times10^{-3}$~\cite{PDG2016}. 
The two-body projections $m(D_{s}^{+}K^{-})$ and $m(K^{+}K^{-})$ are obtained for the signal component using the \sPlot technique, shown in Fig.~\ref{fig:DsKK_Projections}. No significant peak is observed in the \phiz region of the $m(\Kp\Km)$ plot; rather a broad distribution of candidates is found in the region up to $m(\Kp\Km) \simeq 1900 \mevcc$. 

\subsection{Search for \boldmath{$\decay{\Bp}{\Dsp\phiz}$} candidates}

The fit to $\decay{\Bp}{\Dsp\phiz}$ candidates finds a total yield of $N(\decay{\Bp}{\Dsp\phiz}) = 5.3 \pm 6.7$, summed across all categories and \Dsp meson decay modes. 
A yield of $N(\decay{\Bp}{\Dsp\Km\Kp}) = 65 \pm 10 $ is found, consistent with the yield obtained from the full $\decay{\Bp}{\Dsp\Kp\Km}$ measurement. 
The branching fraction for $ \decay{\Bp}{\Dsp\phiz}$ decays is calculated as

\begin{equation}
\mathcal{B}(\decay{\Bp}{\Dsp\phiz}) = R \times \frac{\mathcal{B}(\decay{\Dzb}{\Kp\Km})}{\mathcal{B}(\decay{\phi}{\Kp\Km})} \times \mathcal{B}(\decay{\Bp}{\Dsp\Dzb}),
\label{eq:branching_fraction_calc}
\end{equation}
where the branching fraction $\mathcal{B}(\decay{\phi}{\Kp\Km})= 0.489 \pm 0.005$ has been used~\cite{PDG2016}. 

The free variable $R$ is defined to be the ratio of the signal and normalisation yields, corrected for the selection efficiencies.
The yield of signal candidates in each \Dsp mode is constructed from $R$ and the normalisation yield for the given \Dsp decay mode, $N(\decay{\Bp}{\Dsp\Dzb})$. The product of these two quantities is corrected by the ratio of selection efficiencies

\begin{equation}
N(\decay{\Bp}{\Dsp\phiz}) = R \times N(\decay{\Bp}{\Dsp\Dzb}) \times \frac{\epsilon(\decay{\Bp}{\Dsp\phiz})}{\epsilon(\decay{\Bp}{\Dsp\Dzb})}.
\label{eq:branching_fraction_R}
\end{equation}

The simultaneous fit measures a single value of $R$ for all \Dsp decay mode categories. From an ensemble of pseudoexperiments, $R$ is distributed normally. It can be written as the ratio of signal and normalisation branching fractions using Eq.~{\ref{eq:branching_fraction_calc}. The value is determined to be 

\begin{equation*}
R = \frac{\mathcal{B}(\decay{\Bp}{\Dsp\phiz})}{\mathcal{B}(\decay{\Bp}{\Dsp\Dzb})}\times \frac{\mathcal{B}(\decay{\phi}{\Kp\Km})}{\mathcal{B}(\decay{\Dzb}{\Kp\Km})} =(1.6^{+2.2}_{-1.9}\pm 1.1) \times 10^{-3}, 
\end{equation*}
where the first uncertainty is statistical and the second systematic. This corresponds to a branching fraction for $\decay{\Bp}{\Dsp\phiz}$ decays of

\begin{equation*}
\mathcal{B}(\decay{\Bp}{\Dsp\phiz}) = (1.2^{+1.6}_{-1.4} \pm 0.8  \pm 0.1)\times 10^{-7},
\label{eq:branching_fraction}
\end{equation*}
where the first uncertainty is statistical, the second systematic, and the third results from the uncertainty on the branching fractions $\mathcal{B}(\decay{\Bp}{\Dsp\Dzb})$, $\mathcal{B}(\decay{\phi}{\Kp\Km})$ and $\mathcal{B}(\decay{\Dzb}{\Kp\Km})$. Considering only the statistical uncertainty, the significance of the $\decay{\Bp}{\Dsp\phiz}$ signal is 0.8 standard deviations ($\sigma$). 

Upper limits at 95\% and 90\% confidence levels (CL) are determined using the Feldman-Cousins approach~\cite{FeldmanCousins}. An ensemble of pseudoexperiments is generated for different values of the branching fraction $\mathcal{B}(\decay{\Bp}{\Dsp\phiz})$. These generated pseudoexperiments are then fitted with the nominal fit model to calculate the fitted branching fraction and associated statistical uncertainty, $\sigma_{\text{stat}}$. This method constructs confidence bands based on a likelihood ratio method, calculating the probability of fitting a branching fraction for a given generated branching fraction. This probability is assumed to follow a Gaussian distribution with width $\sigma = \sqrt{\sigma_{\text{stat}}^{2}+\sigma_{\text{syst}}^{2}}$, where $\sigma_{\text{stat}}$ and $\sigma_{\text{syst}}$ are the statistical and systematic uncertainties. The dominant source of systematic uncertainty in this measurement is from the background PDFs. As the size of this uncertainty is not expected to vary as a function of the generated branching fraction, $\sigma_{\text{syst}}$ is assumed to be constant. Nuisance parameters are accounted for using the plug-in method~\cite{plugin}. The generated confidence bands are shown in Fig.~\ref{fig:limit}, where the statistical-only 90\% CL and 95\% CL bands are shown, along with the 95\% CL band with systematic uncertainty included. 
This corresponds to a statistical-only 95\% (90\%) CL limit of $\mathcal{B}(\decay{\Bp}{\Dsp\phiz}) < 4.4 \times 10^{-7}~(3.9 \times 10^{-7})$, and a 95\% (90\%) CL limit including systematic uncertainties of
\begin{equation*}
\mathcal{B}(\decay{\Bp}{\Dsp\phiz}) < 4.9 \times 10^{-7}~(4.2 \times 10^{-7}).
\label{eq:upperlimit}
\end{equation*}
\begin{figure}[t]
    \centering
    \includegraphics[width=0.7\textwidth]{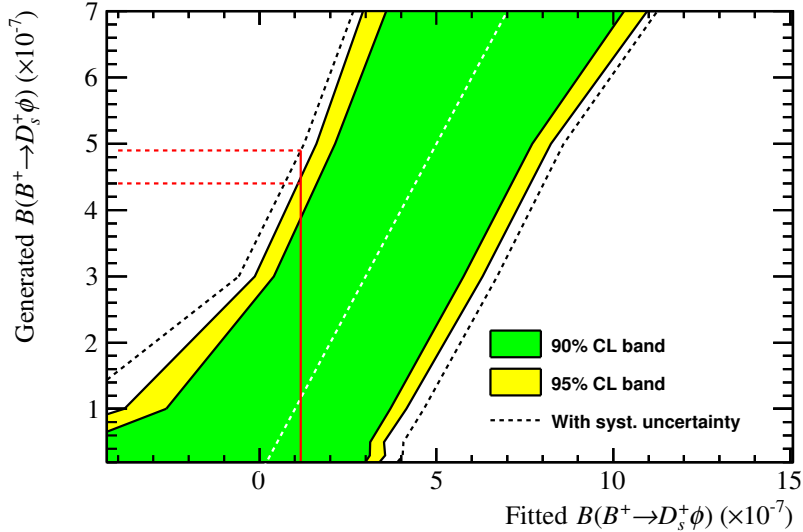}
    
    \caption{Confidence bands produced using the Feldman-Cousins approach. The green and yellow bands represent the statistical-only 90\% and 95\% CL bands. The black dotted line represents the 95\% limit including systematic uncertainties. The measured value of the branching fraction is shown by the vertical red line, and the corresponding 95\% CL limits, with and without systematic uncertainties, are represented by the dotted red lines.}
    \label{fig:limit}   
\end{figure}

\section{Conclusions}

A search for $\decay{\Bp}{\Dsp\Kp\Km}$ decays is performed. The branching fraction is determined to be
\begin{equation*}
\mathcal{B}(\decay{\Bp}{\Dsp\Kp\Km} ) = (7.1 \pm 0.5 \pm 0.6 \pm 0.7) \times 10^{-6}, 
\end{equation*}

\noindent where the first uncertainty is statistical, the second systematic and the third is due to the uncertainty on 
the branching fraction of the normalisation mode $\decay{\Bp}{\Dsp\Dzb}$.
This is the first observation of this decay mode.
A search is also performed for the pure annihilation decay $\decay{\Bp}{\Dsp\phiz}$, but no significant signal is observed and a limit of
\begin{equation*}
\mathcal{B}(\decay{\Bp}{\Dsp\phiz}) < 4.9 \times 10^{-7}~(4.2 \times 10^{-7})
\end{equation*}
is set on the branching fraction at 95\% (90\%) confidence level. The limit on $\mathcal{B}(\decay{\Bp}{\Dsp\phiz})$ presented here supersedes the previous result from \lhcb~\cite{Aaij:2012zh}.

This updated analysis benefits from the significantly larger data sample now available at \lhcb to increase the reach of these searches.
The previous measurement performed by \lhcb reported evidence for the decay $\decay{\Bp}{\Dsp\phiz}$ with a significance greater than 3$\sigma$. 
This update determines that there is a sizeable contribution from $\decay{\Bp}{\Dsp\Kp\Km}$ decays that contribute within the $\phiz$-meson mass window that was previously not considered. 
The result is consistent with the prediction that rescattering contributions to $\decay{\Bp}{\Dsp\phiz}$ decays are small.

%% file: acknowledgements.tex
\section*{Acknowledgements}
%
%
\noindent We express our gratitude to our colleagues in the CERN
accelerator departments for the excellent performance of the LHC. We
thank the technical and administrative staff at the LHCb
institutes. We acknowledge support from CERN and from the national
agencies: CAPES, CNPq, FAPERJ and FINEP (Brazil); MOST and NSFC
(China); CNRS/IN2P3 (France); BMBF, DFG and MPG (Germany); INFN
(Italy); NWO (The Netherlands); MNiSW and NCN (Poland); MEN/IFA
(Romania); MinES and FASO (Russia); MinECo (Spain); SNSF and SER
(Switzerland); NASU (Ukraine); STFC (United Kingdom); NSF (USA).  We
acknowledge the computing resources that are provided by CERN, IN2P3
(France), KIT and DESY (Germany), INFN (Italy), SURF (The
Netherlands), PIC (Spain), GridPP (United Kingdom), RRCKI and Yandex
LLC (Russia), CSCS (Switzerland), IFIN-HH (Romania), CBPF (Brazil),
PL-GRID (Poland) and OSC (USA). We are indebted to the communities
behind the multiple open-source software packages on which we depend.
Individual groups or members have received support from AvH Foundation
(Germany), EPLANET, Marie Sk\l{}odowska-Curie Actions and ERC
(European Union), ANR, Labex P2IO, ENIGMASS and OCEVU, and R\'{e}gion
Auvergne-Rh\^{o}ne-Alpes (France), RFBR and Yandex LLC (Russia), GVA,
XuntaGal and GENCAT (Spain), Herchel Smith Fund, the Royal Society,
the English-Speaking Union and the Leverhulme Trust (United Kingdom).

%% file: LHCb_Authorship_flat_02-Aug-2017.tex
\centerline{\large\bf LHCb collaboration}
\begin{flushleft}
\small
R.~Aaij$^{40}$,
B.~Adeva$^{39}$,
M.~Adinolfi$^{48}$,
Z.~Ajaltouni$^{5}$,
S.~Akar$^{59}$,
J.~Albrecht$^{10}$,
F.~Alessio$^{40}$,
M.~Alexander$^{53}$,
A.~Alfonso~Albero$^{38}$,
S.~Ali$^{43}$
% ...

{\footnotesize \it
% ...
$ ^{5}$Clermont Universit{\'e}, Universit{\'e} Blaise Pascal, CNRS/IN2P3, LPC, Clermont-Ferrand, France\\
% ...
$ ^{10}$Fakult{\"a}t Physik, Technische Universit{\"a}t Dortmund, Dortmund, Germany\\
% ...
$ ^{38}$ICCUB, Universitat de Barcelona, Barcelona, Spain\\
$ ^{39}$Instituto Galego de F{\'\i}sica de Altas Enerx{\'\i}as (IGFAE), Universidade de Santiago de Compostela, Santiago de Compostela, Spain\\
$ ^{40}$European Organization for Nuclear Research (CERN), Geneva, Switzerland\\
% ...
$ ^{43}$Nikhef National Institute for Subatomic Physics, Amsterdam, The Netherlands\\
% ...
$ ^{48}$H.H. Wills Physics Laboratory, University of Bristol, Bristol, United Kingdom\\
% ...
$ ^{53}$School of Physics and Astronomy, University of Glasgow, Glasgow, United Kingdom\\
% ...
$ ^{59}$University of Cincinnati, Cincinnati, OH, United States\\
% ...
}
\end{flushleft}

%% file: main.bbl
\ifx\mcitethebibliography\mciteundefinedmacro
\PackageError{LHCb.bst}{mciteplus.sty has not been loaded}
{This bibstyle requires the use of the mciteplus package.}\fi
\providecommand{\href}[2]{#2}
\begin{mcitethebibliography}{10}
\mciteSetBstSublistMode{n}
\mciteSetBstMaxWidthForm{subitem}{\alph{mcitesubitemcount})}
\mciteSetBstSublistLabelBeginEnd{\mcitemaxwidthsubitemform\space}
{\relax}{\relax}

\bibitem{Aubert:2006xy}
BaBar collaboration, B.~Aubert {\em et~al.},
  \ifthenelse{\boolean{articletitles}}{\emph{{Evidence for the rare decay
  $B^{+} \to D^+_{s} \pi^0$}},
  }{}\href{http://dx.doi.org/10.1103/PhysRevLett.98.171801}{Phys.\ Rev.\ Lett.\
   \textbf{98} (2007) 171801},
  \href{http://arxiv.org/abs/hep-ex/0611030}{{\normalfont\ttfamily
  arXiv:hep-ex/0611030}}\relax
\mciteBstWouldAddEndPuncttrue
\mciteSetBstMidEndSepPunct{\mcitedefaultmidpunct}
{\mcitedefaultendpunct}{\mcitedefaultseppunct}\relax
\EndOfBibitem
\bibitem{Gronau:2012gs}
M.~Gronau, D.~London, and J.~L. Rosner,
  \ifthenelse{\boolean{articletitles}}{\emph{{Rescattering contributions to
  rare B-meson decays}},
  }{}\href{http://dx.doi.org/10.1103/PhysRevD.87.036008}{Phys.\ Rev.\
  \textbf{D87} (2013) 036008},
  \href{http://arxiv.org/abs/1211.5785}{{\normalfont\ttfamily
  arXiv:1211.5785}}\relax
\mciteBstWouldAddEndPuncttrue
\mciteSetBstMidEndSepPunct{\mcitedefaultmidpunct}
{\mcitedefaultendpunct}{\mcitedefaultseppunct}\relax
\EndOfBibitem
\bibitem{Zou:2009zza}
H.~Zou, R.-H. Li, X.-X. Wang, and C.-D. Lu,
  \ifthenelse{\boolean{articletitles}}{\emph{{The CKM suppressed
  $\decay{B(B_s)}{\Db_{(s)}P, \Db_{(s)}V,\Db^{*}_{(s)}P, \Db^{*}_{(s)}V}$
  decays in perturbative QCD approach}},
  }{}\href{http://dx.doi.org/10.1088/0954-3899/37/1/015002}{J.\ Phys.\
  \textbf{G37} (2010) 015002},
  \href{http://arxiv.org/abs/0908.1856}{{\normalfont\ttfamily
  arXiv:0908.1856}}\relax
\mciteBstWouldAddEndPuncttrue
\mciteSetBstMidEndSepPunct{\mcitedefaultmidpunct}
{\mcitedefaultendpunct}{\mcitedefaultseppunct}\relax
\EndOfBibitem
\bibitem{Mohanta:2002wf}
R.~Mohanta, \ifthenelse{\boolean{articletitles}}{\emph{{Searching for new
  physics in the rare decay $B^{+} \to D_s^+ \phi$}},
  }{}\href{http://dx.doi.org/10.1016/S0370-2693(02)02173-1}{Phys.\ Lett.\
  \textbf{B540} (2002) 241},
  \href{http://arxiv.org/abs/hep-ph/0205297}{{\normalfont\ttfamily
  arXiv:hep-ph/0205297}}\relax
\mciteBstWouldAddEndPuncttrue
\mciteSetBstMidEndSepPunct{\mcitedefaultmidpunct}
{\mcitedefaultendpunct}{\mcitedefaultseppunct}\relax
\EndOfBibitem
\bibitem{Mohanta:2007uu}
R.~Mohanta and A.~K. Giri, \ifthenelse{\boolean{articletitles}}{\emph{{Possible
  signatures of unparticles in rare annihilation type B decays}},
  }{}\href{http://dx.doi.org/10.1103/PhysRevD.76.057701}{Phys.\ Rev.\
  \textbf{D76} (2007) 057701},
  \href{http://arxiv.org/abs/0707.3308}{{\normalfont\ttfamily
  arXiv:0707.3308}}\relax
\mciteBstWouldAddEndPuncttrue
\mciteSetBstMidEndSepPunct{\mcitedefaultmidpunct}
{\mcitedefaultendpunct}{\mcitedefaultseppunct}\relax
\EndOfBibitem
\bibitem{Lu:2001yz}
C.-D. Lu, \ifthenelse{\boolean{articletitles}}{\emph{{Calculation of pure
  annihilation type decay \decay{\Bp}{\Dsp\phiz}}},
  }{}\href{http://dx.doi.org/10.1007/s100520200929}{Eur.\ Phys.\ J.\
  \textbf{C24} (2002) 121},
  \href{http://arxiv.org/abs/hep-ph/0112127}{{\normalfont\ttfamily
  arXiv:hep-ph/0112127}}\relax
\mciteBstWouldAddEndPuncttrue
\mciteSetBstMidEndSepPunct{\mcitedefaultmidpunct}
{\mcitedefaultendpunct}{\mcitedefaultseppunct}\relax
\EndOfBibitem
\bibitem{fB:2013HPQCD}
HPQCD collaboration, R.~J. Dowdall {\em et~al.},
  \ifthenelse{\boolean{articletitles}}{\emph{{$B$-meson decay constants from
  improved lattice nonrelativistic QCD and physical $u$, $d$, $s$ and $c$
  quarks}}, }{}\href{http://dx.doi.org/10.1103/PhysRevLett.110.222003}{Phys.\
  Rev.\ Lett.\  \textbf{110} (2013) 222003},
  \href{http://arxiv.org/abs/1302.2644}{{\normalfont\ttfamily
  arXiv:1302.2644}}\relax
\mciteBstWouldAddEndPuncttrue
\mciteSetBstMidEndSepPunct{\mcitedefaultmidpunct}
{\mcitedefaultendpunct}{\mcitedefaultseppunct}\relax
\EndOfBibitem
\bibitem{fB:2016ETM}
ETM collaboration, A.~Bussone {\em et~al.},
  \ifthenelse{\boolean{articletitles}}{\emph{{Mass of the $b$ quark and
  $B$-meson decay constants from ${N}_{f}=2+1+1$ twisted-mass lattice QCD}},
  }{}\href{http://dx.doi.org/10.1103/PhysRevD.93.114505}{Phys.\ Rev.\
  \textbf{D93} (2016) 114505},
  \href{http://arxiv.org/abs/1603.04306}{{\normalfont\ttfamily
  arXiv:1603.04306}}\relax
\mciteBstWouldAddEndPuncttrue
\mciteSetBstMidEndSepPunct{\mcitedefaultmidpunct}
{\mcitedefaultendpunct}{\mcitedefaultseppunct}\relax
\EndOfBibitem
\bibitem{fB:2016Fermi}
Fermilab Lattice and MILC collaborations, A.~Bazavov {\em et~al.},
  \ifthenelse{\boolean{articletitles}}{\emph{{$B^{0}_{(s)}$-mixing matrix
  elements from lattice QCD for the Standard Model and beyond}},
  }{}\href{http://dx.doi.org/10.1103/PhysRevD.93.113016}{Phys.\ Rev.\
  \textbf{D93} (2016) 113016},
  \href{http://arxiv.org/abs/1602.03560}{{\normalfont\ttfamily
  arXiv:1602.03560}}\relax
\mciteBstWouldAddEndPuncttrue
\mciteSetBstMidEndSepPunct{\mcitedefaultmidpunct}
{\mcitedefaultendpunct}{\mcitedefaultseppunct}\relax
\EndOfBibitem
\bibitem{Aaij:2012zh}
LHCb collaboration, R.~Aaij {\em et~al.},
  \ifthenelse{\boolean{articletitles}}{\emph{{First evidence for the
  annihilation decay mode $\Bp\to\Dsp\phiz$}},
  }{}\href{http://dx.doi.org/10.1007/JHEP02(2013)043}{JHEP \textbf{02} (2013)
  043}, \href{http://arxiv.org/abs/1210.1089}{{\normalfont\ttfamily
  arXiv:1210.1089}}\relax
\mciteBstWouldAddEndPuncttrue
\mciteSetBstMidEndSepPunct{\mcitedefaultmidpunct}
{\mcitedefaultendpunct}{\mcitedefaultseppunct}\relax
\EndOfBibitem
\bibitem{PDG2016}
Particle Data Group, C.~Patrignani {\em et~al.},
  \ifthenelse{\boolean{articletitles}}{\emph{{\href{http://pdg.lbl.gov/}{Review
  of particle physics}}},
  }{}\href{http://dx.doi.org/10.1088/1674-1137/40/10/100001}{Chin.\ Phys.\
  \textbf{C40} (2016) 100001}\relax
\mciteBstWouldAddEndPuncttrue
\mciteSetBstMidEndSepPunct{\mcitedefaultmidpunct}
{\mcitedefaultendpunct}{\mcitedefaultseppunct}\relax
\EndOfBibitem
\bibitem{Alves:2008zz}
LHCb collaboration, A.~A. Alves~Jr.\ {\em et~al.},
  \ifthenelse{\boolean{articletitles}}{\emph{{The \lhcb detector at the LHC}},
  }{}\href{http://dx.doi.org/10.1088/1748-0221/3/08/S08005}{JINST \textbf{3}
  (2008) S08005}\relax
\mciteBstWouldAddEndPuncttrue
\mciteSetBstMidEndSepPunct{\mcitedefaultmidpunct}
{\mcitedefaultendpunct}{\mcitedefaultseppunct}\relax
\EndOfBibitem
\bibitem{LHCb-DP-2014-002}
LHCb collaboration, R.~Aaij {\em et~al.},
  \ifthenelse{\boolean{articletitles}}{\emph{{LHCb detector performance}},
  }{}\href{http://dx.doi.org/10.1142/S0217751X15300227}{Int.\ J.\ Mod.\ Phys.\
  \textbf{A30} (2015) 1530022},
  \href{http://arxiv.org/abs/1412.6352}{{\normalfont\ttfamily
  arXiv:1412.6352}}\relax
\mciteBstWouldAddEndPuncttrue
\mciteSetBstMidEndSepPunct{\mcitedefaultmidpunct}
{\mcitedefaultendpunct}{\mcitedefaultseppunct}\relax
\EndOfBibitem
\bibitem{BBDT}
V.~V. Gligorov and M.~Williams,
  \ifthenelse{\boolean{articletitles}}{\emph{{Efficient, reliable and fast
  high-level triggering using a bonsai boosted decision tree}},
  }{}\href{http://dx.doi.org/10.1088/1748-0221/8/02/P02013}{JINST \textbf{8}
  (2013) P02013}, \href{http://arxiv.org/abs/1210.6861}{{\normalfont\ttfamily
  arXiv:1210.6861}}\relax
\mciteBstWouldAddEndPuncttrue
\mciteSetBstMidEndSepPunct{\mcitedefaultmidpunct}
{\mcitedefaultendpunct}{\mcitedefaultseppunct}\relax
\EndOfBibitem
\bibitem{Sjostrand:2007gs}
T.~Sj\"{o}strand, S.~Mrenna, and P.~Skands,
  \ifthenelse{\boolean{articletitles}}{\emph{{A brief introduction to PYTHIA
  8.1}}, }{}\href{http://dx.doi.org/10.1016/j.cpc.2008.01.036}{Comput.\ Phys.\
  Commun.\  \textbf{178} (2008) 852},
  \href{http://arxiv.org/abs/0710.3820}{{\normalfont\ttfamily
  arXiv:0710.3820}}\relax
\mciteBstWouldAddEndPuncttrue
\mciteSetBstMidEndSepPunct{\mcitedefaultmidpunct}
{\mcitedefaultendpunct}{\mcitedefaultseppunct}\relax
\EndOfBibitem
\bibitem{Sjostrand:2006za}
T.~Sj\"{o}strand, S.~Mrenna, and P.~Skands,
  \ifthenelse{\boolean{articletitles}}{\emph{{PYTHIA 6.4 physics and manual}},
  }{}\href{http://dx.doi.org/10.1088/1126-6708/2006/05/026}{JHEP \textbf{05}
  (2006) 026}, \href{http://arxiv.org/abs/hep-ph/0603175}{{\normalfont\ttfamily
  arXiv:hep-ph/0603175}}\relax
\mciteBstWouldAddEndPuncttrue
\mciteSetBstMidEndSepPunct{\mcitedefaultmidpunct}
{\mcitedefaultendpunct}{\mcitedefaultseppunct}\relax
\EndOfBibitem
\bibitem{LHCb-PROC-2010-056}
I.~Belyaev {\em et~al.}, \ifthenelse{\boolean{articletitles}}{\emph{{Handling
  of the generation of primary events in Gauss, the LHCb simulation
  framework}}, }{}\href{http://dx.doi.org/10.1088/1742-6596/331/3/032047}{{J.\
  Phys.\ Conf.\ Ser.\ } \textbf{331} (2011) 032047}\relax
\mciteBstWouldAddEndPuncttrue
\mciteSetBstMidEndSepPunct{\mcitedefaultmidpunct}
{\mcitedefaultendpunct}{\mcitedefaultseppunct}\relax
\EndOfBibitem
\bibitem{Lange:2001uf}
D.~J. Lange, \ifthenelse{\boolean{articletitles}}{\emph{{The EvtGen particle
  decay simulation package}},
  }{}\href{http://dx.doi.org/10.1016/S0168-9002(01)00089-4}{Nucl.\ Instrum.\
  Meth.\  \textbf{A462} (2001) 152}\relax
\mciteBstWouldAddEndPuncttrue
\mciteSetBstMidEndSepPunct{\mcitedefaultmidpunct}
{\mcitedefaultendpunct}{\mcitedefaultseppunct}\relax
\EndOfBibitem
\bibitem{Golonka:2005pn}
P.~Golonka and Z.~Was, \ifthenelse{\boolean{articletitles}}{\emph{{PHOTOS Monte
  Carlo: A precision tool for QED corrections in $Z$ and $W$ decays}},
  }{}\href{http://dx.doi.org/10.1140/epjc/s2005-02396-4}{Eur.\ Phys.\ J.\
  \textbf{C45} (2006) 97},
  \href{http://arxiv.org/abs/hep-ph/0506026}{{\normalfont\ttfamily
  arXiv:hep-ph/0506026}}\relax
\mciteBstWouldAddEndPuncttrue
\mciteSetBstMidEndSepPunct{\mcitedefaultmidpunct}
{\mcitedefaultendpunct}{\mcitedefaultseppunct}\relax
\EndOfBibitem
\bibitem{Allison:2006ve}
Geant4 collaboration, J.~Allison {\em et~al.},
  \ifthenelse{\boolean{articletitles}}{\emph{{Geant4 developments and
  applications}}, }{}\href{http://dx.doi.org/10.1109/TNS.2006.869826}{IEEE
  Trans.\ Nucl.\ Sci.\  \textbf{53} (2006) 270}\relax
\mciteBstWouldAddEndPuncttrue
\mciteSetBstMidEndSepPunct{\mcitedefaultmidpunct}
{\mcitedefaultendpunct}{\mcitedefaultseppunct}\relax
\EndOfBibitem
\bibitem{Agostinelli:2002hh}
Geant4 collaboration, S.~Agostinelli {\em et~al.},
  \ifthenelse{\boolean{articletitles}}{\emph{{Geant4: A simulation toolkit}},
  }{}\href{http://dx.doi.org/10.1016/S0168-9002(03)01368-8}{Nucl.\ Instrum.\
  Meth.\  \textbf{A506} (2003) 250}\relax
\mciteBstWouldAddEndPuncttrue
\mciteSetBstMidEndSepPunct{\mcitedefaultmidpunct}
{\mcitedefaultendpunct}{\mcitedefaultseppunct}\relax
\EndOfBibitem
\bibitem{LHCb-PROC-2011-006}
M.~Clemencic {\em et~al.}, \ifthenelse{\boolean{articletitles}}{\emph{{The
  \lhcb simulation application, Gauss: Design, evolution and experience}},
  }{}\href{http://dx.doi.org/10.1088/1742-6596/331/3/032023}{{J.\ Phys.\ Conf.\
  Ser.\ } \textbf{331} (2011) 032023}\relax
\mciteBstWouldAddEndPuncttrue
\mciteSetBstMidEndSepPunct{\mcitedefaultmidpunct}
{\mcitedefaultendpunct}{\mcitedefaultseppunct}\relax
\EndOfBibitem
\bibitem{LHCb-PAPER-2012-050}
LHCb collaboration, R.~Aaij {\em et~al.},
  \ifthenelse{\boolean{articletitles}}{\emph{{First observations of
  $\Bsb\to\Dp\Dm$, $\Dsp\Dm$ and $\Dz\Dzb$ decays}},
  }{}\href{http://dx.doi.org/10.1103/PhysRevD.87.092007}{Phys.\ Rev.\
  \textbf{D87} (2013) 092007},
  \href{http://arxiv.org/abs/1302.5854}{{\normalfont\ttfamily
  arXiv:1302.5854}}\relax
\mciteBstWouldAddEndPuncttrue
\mciteSetBstMidEndSepPunct{\mcitedefaultmidpunct}
{\mcitedefaultendpunct}{\mcitedefaultseppunct}\relax
\EndOfBibitem
\bibitem{Pivk:2004ty}
M.~Pivk and F.~R. Le~Diberder,
  \ifthenelse{\boolean{articletitles}}{\emph{{sPlot: A statistical tool to
  unfold data distributions}},
  }{}\href{http://dx.doi.org/10.1016/j.nima.2005.08.106}{Nucl.\ Instrum.\
  Meth.\  \textbf{A555} (2005) 356},
  \href{http://arxiv.org/abs/physics/0402083}{{\normalfont\ttfamily
  arXiv:physics/0402083}}\relax
\mciteBstWouldAddEndPuncttrue
\mciteSetBstMidEndSepPunct{\mcitedefaultmidpunct}
{\mcitedefaultendpunct}{\mcitedefaultseppunct}\relax
\EndOfBibitem
\bibitem{Breiman}
L.~Breiman, J.~H. Friedman, R.~A. Olshen, and C.~J. Stone, {\em Classification
  and regression trees}, Wadsworth international group, Belmont, California,
  USA, 1984\relax
\mciteBstWouldAddEndPuncttrue
\mciteSetBstMidEndSepPunct{\mcitedefaultmidpunct}
{\mcitedefaultendpunct}{\mcitedefaultseppunct}\relax
\EndOfBibitem
\bibitem{Punzi:2003bu}
G.~Punzi, \ifthenelse{\boolean{articletitles}}{\emph{{Sensitivity of searches
  for new signals and its optimization}}, }{} in {\em Statistical Problems in
  Particle Physics, Astrophysics, and Cosmology} (L.~{Lyons}, R.~{Mount}, and
  R.~{Reitmeyer}, eds.), p.~79, 2003.
\newblock \href{http://arxiv.org/abs/physics/0308063}{{\normalfont\ttfamily
  arXiv:physics/0308063}}\relax
\mciteBstWouldAddEndPuncttrue
\mciteSetBstMidEndSepPunct{\mcitedefaultmidpunct}
{\mcitedefaultendpunct}{\mcitedefaultseppunct}\relax
\EndOfBibitem
\bibitem{Hulsbergen:2005pu}
W.~D. Hulsbergen, \ifthenelse{\boolean{articletitles}}{\emph{{Decay chain
  fitting with a Kalman filter}},
  }{}\href{http://dx.doi.org/10.1016/j.nima.2005.06.078}{Nucl.\ Instrum.\
  Meth.\  \textbf{A552} (2005) 566},
  \href{http://arxiv.org/abs/physics/0503191}{{\normalfont\ttfamily
  arXiv:physics/0503191}}\relax
\mciteBstWouldAddEndPuncttrue
\mciteSetBstMidEndSepPunct{\mcitedefaultmidpunct}
{\mcitedefaultendpunct}{\mcitedefaultseppunct}\relax
\EndOfBibitem
\bibitem{Skwarnicki:1986xj}
T.~Skwarnicki, {\em {A study of the radiative cascade transitions between the
  Upsilon-prime and Upsilon resonances}}, PhD thesis, Institute of Nuclear
  Physics, Krakow, 1986,
  {\href{http://inspirehep.net/record/230779/}{DESY-F31-86-02}}\relax
\mciteBstWouldAddEndPuncttrue
\mciteSetBstMidEndSepPunct{\mcitedefaultmidpunct}
{\mcitedefaultendpunct}{\mcitedefaultseppunct}\relax
\EndOfBibitem
\bibitem{LHCb-PAPER-2017-021}
LHCb collaboration, R.~Aaij {\em et~al.},
  \ifthenelse{\boolean{articletitles}}{\emph{{Measurement of $C\!P$ observables
  in $B^\pm \to D^{(\ast)}K^\pm$ and $B^\pm \to D^{(\ast)}\pi^\pm$ decays}},
  }{}\href{http://arxiv.org/abs/1708.06370}{{\normalfont\ttfamily
  arXiv:1708.06370}}, {submitted to Phys. Lett. B}\relax
\mciteBstWouldAddEndPuncttrue
\mciteSetBstMidEndSepPunct{\mcitedefaultmidpunct}
{\mcitedefaultendpunct}{\mcitedefaultseppunct}\relax
\EndOfBibitem
\bibitem{LHCb-PAPER-2016-006}
LHCb collaboration, R.~Aaij {\em et~al.},
  \ifthenelse{\boolean{articletitles}}{\emph{{Model-independent measurement of
  the CKM angle $\gamma$ using $\Bz \to \D\Kstarz$ decays with
  $\D\to\KS\pip\pim$ and $\KS\Kp\Km$}},
  }{}\href{http://dx.doi.org/10.1007/JHEP06(2016)131}{JHEP \textbf{06} (2016)
  131}, \href{http://arxiv.org/abs/1604.01525}{{\normalfont\ttfamily
  arXiv:1604.01525}}\relax
\mciteBstWouldAddEndPuncttrue
\mciteSetBstMidEndSepPunct{\mcitedefaultmidpunct}
{\mcitedefaultendpunct}{\mcitedefaultseppunct}\relax
\EndOfBibitem
\bibitem{Cranmer:2000du}
K.~S. Cranmer, \ifthenelse{\boolean{articletitles}}{\emph{{Kernel estimation in
  high-energy physics}},
  }{}\href{http://dx.doi.org/10.1016/S0010-4655(00)00243-5}{Comput.\ Phys.\
  Commun.\  \textbf{136} (2001) 198},
  \href{http://arxiv.org/abs/hep-ex/0011057}{{\normalfont\ttfamily
  arXiv:hep-ex/0011057}}\relax
\mciteBstWouldAddEndPuncttrue
\mciteSetBstMidEndSepPunct{\mcitedefaultmidpunct}
{\mcitedefaultendpunct}{\mcitedefaultseppunct}\relax
\EndOfBibitem
\bibitem{FeldmanCousins}
G.~J. Feldman and R.~D. Cousins, \ifthenelse{\boolean{articletitles}}{\emph{A
  unified approach to the classical statistical analysis of small signals},
  }{}\href{http://dx.doi.org/10.1103/PhysRevD.57.3873}{Phys.\ Rev.\
  \textbf{D57} (1998) 3873},
  \href{http://arxiv.org/abs/physics/9711021}{{\normalfont\ttfamily
  arXiv:physics/9711021}}\relax
\mciteBstWouldAddEndPuncttrue
\mciteSetBstMidEndSepPunct{\mcitedefaultmidpunct}
{\mcitedefaultendpunct}{\mcitedefaultseppunct}\relax
\EndOfBibitem
\bibitem{plugin}
B.~Sen, M.~Walker, and M.~Woodroofe,
  \ifthenelse{\boolean{articletitles}}{\emph{On the unified method with
  nuisance parameters}, }{}Statistica Sinica \textbf{19} (2009) 301\relax
\mciteBstWouldAddEndPuncttrue
\mciteSetBstMidEndSepPunct{\mcitedefaultmidpunct}
{\mcitedefaultendpunct}{\mcitedefaultseppunct}\relax
\EndOfBibitem
\end{mcitethebibliography}
